# Sharpening Shapley Allocation: from Basel 2.5 to FRTB

**Marco Scaringi[1,2], Marco Bianchetti[1,3,*]**

[1] Financial and Market Risk Management, Intesa Sanpaolo, Milan, Italy
[2] Currently at Trading Quant Department, ING Bank N. V., Milan, Italy
[3] Department of Statistical Sciences "Paolo Fortunati", University of Bologna, Italy
* Corresponding author, marco.bianchetti@unibo.it

## Abstract

Risk allocation, the decomposition of a portfolio-wide risk measure into component contributions, is a fundamental problem in financial risk management due to the non-additive nature of risk measures, the layered organizational structures of financial institutions, and the range of possible allocation strategies characterized by different rationales and properties.
In this work, we conduct a systematic review of the major risk allocation strategies typically used in finance, comparing their theoretical properties, practical advantages, and limitations. To this scope we set up a specific testing framework, including both simplified settings, designed to highlight basic intrinsic behaviours, and realistic financial portfolios under different risk regulations, i.e. Basel 2.5 and FRTB. Furthermore, we develop and test novel practical solutions to manage the issue of negative risk allocations and of multi-level risk allocation in the layered organizational structure of financial institutions, while preserving the additivity property. Finally, we devote particular attention to the computational aspects of risk allocation.
Our results show that, in this context, the Shapley allocation strategy offers the best compromise between simplicity, mathematical properties, risk representation and computational cost. The latter is still acceptable even in the challenging case of many business units, provided that an efficient Monte Carlo simulation is employed, which offers excellent scaling and convergence properties. While our empirical applications focus on market risk, our methodological framework is fully general and applicable to other financial context such as valuation risk, liquidity risk, credit risk, and counterparty credit risk.

**Classifications**: **JEL**: C63, D81, G11, G28, G32; **MSC**: 91B30 (primary), 91A12, 91G70, 62P05, 65C05 (secondary); **ACM**: G.3, I.2.11, J.4, J.1.

### Acknowledgements

The authors gratefully acknowledge fruitful interactions with F. Lania, L. Lamorte, D. Maffei, G. Polito, and many other colleagues at Intesa Sanpaolo financial risk management and front office trading desks, and with C. Acerbi, F. Centrone, G. Fusai, E. R. Gianin, U. Cherubini.

# Summary

# 1. Introduction

## 1.1. Risk Allocation in Finance

To protect themselves from the uncertainty about their future net worth, financial institutions subject to regulatory requirements must hold an amount of riskless assets, called *risk capital*. Risk capital is typically related to a combination of risk measures associated to the different risk sources (credit, market, liquidity, operational, etc.). After the 2008 financial crisis, regulators worldwide strengthened capital requirements, forcing institutions to raise more capital to support their business. Therefore, institutions were forced to reallocate internal resources taking into account not only expected profits but also capital consumption of business units.

The problem of *risk aggregation* consists in combining different risk measures into the total economic capital of the institution. The other way round, the problem of *risk allocation* (or *risk attribution*) consists into the decomposition of a risk measure into the contributions of its components (e.g. each single trade, or sub-portfolio, business unit, underlying risk factor, etc.), contributing to the measure itself. A typical example is the internal market risk reporting in a bank: the different business units are organised according to some *hierarchy* and manage portfolios of financial instruments (bonds, derivatives, funds, loans, etc.), while the market risk management unit produces the market risk figures (e.g. VaR, stressed VaR, Expected Shortfall, etc.), monitors the corresponding risk limits and reports to business units and management. The risk aggregation/allocation problem is not straightforward since many risk measures are typically non-additive, i.e. the overall portfolio risk measure, which determines the risk capital, is typically (but not necessarily) smaller than the sum of risk measures at business unit level due to interactions and correlations among its risk components (i.e. because of diversification and hedging effects). Thus, on the one hand, the single (portfolio level) risk measures are not representative of the actual capital consumption of each trading desk, while, on the other hand, it’s not trivial how to split the total capital figure and allocate it to each business unit. We report in App. 8.1 a (non-exhaustive) list of risk measures which are typically encountered in financial risk management.

From a financial and risk management point of view, a transparent, statistically robust and performant risk allocation strategy clearly provides several benefits. Firstly, it offers a clear and reliable representation of risk sources at a given aggregation level considering correlations, diversification and hedging effects, enhancing the overall understanding of the risk map, and improving transparency and accuracy of risk reporting. Moreover, it facilitates what-if analyses, i.e. simulations of possible portfolio changes, e.g. portfolio merging or split, impact of new trades, etc. Finally, it allows to spot possible risk-based business opportunities, e.g. trading or hedging strategies with favourable P&L/capital ratio, and a better performance management of business units using appropriate KPIs which combine P&L and capital consumption.

From a mathematical point of view, risk allocation is a challenging task which presents several complexities. The main challenge lies in the non-additivity of popular risk measures, which makes standalone allocation, the simplest strategy, not satisfactory. Furthermore, risk measures are typically based on empirical realizations of profits and losses (P&L), using historical or Monte Carlo approaches. Hence, viable risk allocation strategies must be general must be general and avoid strong mathematical assumptions (unless some approximations are considered) and must be robust against possible statistical instabilities associated with tail measures, i.e. when one samples P&L distributions tails including only a few scenarios.

Finally, from a computational and IT point of view, risk measures calculation processes may be challenging, e.g. in case of Monte Carlo simulation, leading to computational bottlenecks for those allocation strategies which require several risk measure recalculations. Furthermore, risk allocation at lowest hierarchical levels (i.e. single trade, or single risk scenario, or single risk factor, etc.) requires to handle and store huge amount of data. Therefore, the risk allocation task requires a performant and flexible infrastructure able to integrate all the required components: trades, portfolio hierarchy, risk factors, risk scenarios, stress test, what-if analyses, pricing and risk analytics, data storage, hardware, etc.

In this paper we focus on the *Shapley allocation strategy*, which, according to our results based on the comparison with other alternative allocation strategies, provides the best compromise between financial, mathematical and computational properties.

## 1.2. Literature Review

There is an extensive scientific literature related to risk allocation in general and to Shapley allocation in particular. Shapley allocation was originally developed by Lloyd Shapley[1] in 1953 (see Shapley (1953), Roth (1988)) in the context of cooperative game theory. The basic idea is that a set of players may form different coalitions to individually optimize a common worth function. To achieve this result, each player is allocated a fair worth that corresponds to its average marginal contribution to the worth of all possible coalitions. Currently, Shapley allocation is widely used in many different fields, adapting the concepts of players, coalitions and worth function to the specific problem, as briefly discussed below.

In the field of *artificial intelligence*, Li et al. (2024) and Rozemberczki et al. (2022) provided a detailed review of Shapley allocation applications in machine learning for features importance and explainable artificial intelligence (XAI). Brigo et al. (2021) focused on neural networks calibration of the popular Heston stochastic volatility model for option pricing, showing that Shapley allocation may help to choose the network architecture and to explain the importance of Heston model parameters. Aiolfi et al. (2024) (see their fig. 5) used feature importance to confirm that Bermudan Swaption prices are principally driven by their corresponding maximum underlying European Swaptions.

In the field of *financial portfolio optimization*, Mussard and Terraza (2008) used the Shapley allocation to decompose the portfolio covariance matrix and to define new variance ratios which measure the contribution of each security to the overall portfolio risk. Ortmann (2016) found a relationship between the Shapley allocation and the beta factor in the Capital Asset Pricing Model (CAPM). Moehle et al. (2021) used Shapley allocation to attribute portfolio performance to different features driving the investment process. Shalit (2023) used the weighted Shapley allocation, admitting players' asymmetries, to attribute risk and return of an efficient mean-variance portfolio to portfolio's components.

In the field of *financial risk aggregation*, Morone et al. (2007) developed independent marginal models for the loss distributions of risk classes and merge them into a joint distribution via copula. Successively, Lugoboni et al. (2021) improved this approach using Shapley allocation.

In the field of *financial risk allocation*, Denault (2001) took an axiomatic approach suggesting a set of properties to be fulfilled by any coherent risk allocation strategy connected to coherent risk measures. Epperlein and Smillie (2006) used kernel estimators to compute component VaR (i.e. VaR allocation to sub-portfolios). Tasche (2008) examined in detail the Euler allocation deriving specific formulas for some popular risk measures. Li et al (2016) proposed a Constrained Aumann-Shapley (CAS) allocation strategy at single trade level which preserves additivity and associativity properties at any hierarchical level. Balog et al (2017) systematically analysed 7 different allocation strategies in terms of ten reasonable fairness properties. Centrone and Rosazza Gianin (2018) studied capital allocation rules satisfying suitable properties for convex and quasi-convex risk measures, by focusing on Aumann–Shapley allocation. Hagan et al. (2021) used Shapley allocation strategy to allocate risk to non-orthogonal greeks in a portfolio of derivatives, based on previous results of Colini-Baldeschi et al. (2018) on variance and volatility games, where the risk measure is the variance or the volatility of a portfolio. Mastrogiacomo and Rosazza Gianin (2024) studied risk allocation for dynamic risk measures with an axiomatic approach and by exploiting the relation between risk measures and BSDEs. Holden (2024) discussed capital allocation using the Euler formula and Value-at-Risk (VaR) and Expected shortfall (ES).

In the specific context of the *Fundamental Review of the Trading Book* (FRTB), Schulze (2018) presented a way to compute Euler allocation for the six risk measures contributing to the FRTB capital charge. Li and Xing (2019) focused on the Internal Model Approach (IMA) under FRTB and proposed two computationally efficient Euler allocation strategies weighted according to FRTB liquidity horizons.

[1] Awarded with the Nobel prize in Economic Sciences in 2012, https://www.nobelprize.org/prizes/economic-sciences/2012/shapley/facts/.

Regarding Monte Carlo simulation, introduced by Mann and Shapley (1960), Mitchell et al. (2022) exploited efficient estimation of Shapley allocation using different variance reduction techniques. Markov chain Monte Carlo was used by Holden (2024).

## 1.3. Our Contribution

A key contribution of our study is a detailed evaluation of the most common risk allocation strategies used in finance and risk management, with particular attention to both their theoretical properties and their scalability and practical feasibility in large-scale financial applications. Specifically, we prove that the Shapley allocation strategy remains a feasible choice even in the challenging case of tens of business units, provided that, once a certain number of business units is exceeded, one switches from the analytical allocation (which scales exponentially with the number of business units) to the Monte Carlo allocation (which scales linearly with the number of business units), thanks to its excellent convergence properties.

Moreover, we develop innovative solutions to address the challenge of negative risk allocations. While prior literature recognized the occurrence of negative risk contributions, particularly in Shapley-based allocations, no structured or widely accepted solutions have been established. Although negative values are theoretically valid outcomes of an allocation strategy, they can be problematic in practical applications where non-negative risk contributions are often required for interpretability. To address this gap, we conduct a rigorous analysis of the theoretical conditions under which negative allocations may arise, and we propose some novel practical approaches that preserve the fundamental principles of Shapley allocation while ensuring non-negative outcomes.

Another contribution of our study is the development of a multi-level Shapley allocation strategy. Unlike other approaches, which typically focus on a single hierarchical level, or prevent marginal contributions from different hierarchical branches (see e.g. Li et al. (2016)), our framework systematically accounts for a consistent risk allocation which preserves the additivity property at multiple levels, a typical situation of large financial institutions where business units are organized in a deep hierarchical structure. To enhance flexibility, we propose different viable multi-level allocation strategies, each based on a different rationale. The choice between these alternatives may be guided by the specific objectives of the risk allocation exercise and by internal managerial decisions.

Finally, we set up a specific testing framework, including both simplified settings, designed to highlight basic behaviours of allocation strategies of different risk measures, and realistic financial portfolios, which reflect the typical trading activities of a financial institution, providing detailed numerical analyses in two market risk contexts of particular interest: Basel 2.5 (see CRR1 (2013) and CRR2 (2019)) and Fundamental Review of the Trading Book (FRTB, see CRR3 (2024)) regulatory frameworks.

Overall, our results show that, even in the context of complex financial institutions, characterized by a multi-level organisational structure including many business units, which make use of regulatory risk measures lacking desirable theoretical properties (e.g. historical VaR), the Shapley allocation strategy offers the best compromise between simplicity, mathematical properties, risk representation and computational cost. While the empirical applications of this paper focus on market risk, our methodological framework is fully general and applicable, in principle, to any financial context where a global risk measure must be decomposed into the contributions of individual components, including - among others - valuation risk, liquidity risk, credit risk, and counterparty credit risk.

The structure of the paper is the following: in Sec. 2 we introduce the notation and describe the methodological framework; in Sec. 3 we discuss some numerical methods used to speed up the calculation of allocation strategies; in Sec. 4 we test the behaviour of the different allocation strategies in a simplified context; in Sec. 5 we provide numerical results in the real world of regulatory capital allocation; in Sec. 6 we summarize our conclusions.

# 2. Methodology

In this section we set the theoretical framework used throughout the paper, aiming to provide a comprehensive overview for readers unfamiliar with the subject. Experienced readers can skip to the next sections. In Sec. 2.1 we remind the concepts, language and notation of *game theory*, which provides an excellent framework to formalize and discuss the risk allocation problem. In Sec. 2.2 we translate the concepts from game theory to risk allocation theory. In Sec. 2.3 we summarize and compare several popular allocation strategies, with a particular focus on Shapley allocation and its Monte Carlo simulation.

## 2.1. Game Theory

We consider a *game* with a set of $n$ players $\mathcal{P} = \{p_1, \dots, p_n\}$ and we denote by $\mathcal{N} = \{1, \dots, n\}$ the set of indices labelling the elements of $\mathcal{P}$. Since there is a one-to-one correspondence between $\mathcal{P}$ and $\mathcal{N}$, we use directly $\mathcal{N}$ and forget about $\mathcal{P}$. Any subset of players $\mathcal{S} \subseteq \mathcal{N}$ is called a *coalition*, and $\mathcal{N}$ itself is called the *grand coalition*. We assume that all players are equal (i.e. subject to the same rules, no asymmetries) and behave independently of each other.

We define the game *characteristic function* (a.k.a. *value, utility, cost, contribution, reward* or *objective function*) $v: P(\mathcal{N}) \to \mathbb{R}$, where $P(\mathcal{N})$ is the power set of $\mathcal{N}$ (i.e. the set of all subsets of $\mathcal{N}$ with cardinality $|P(\mathcal{N})| = 2^n$), such that $v(\{i\}), v(\mathcal{S}), v(\mathcal{N}),$ represent the value that, respectively, i-th player, coalition $\mathcal{S}$, grand coalition $\mathcal{N}$ may obtain acting by themselves, i.e. respectively, alone, coalized with the other players in $\mathcal{S}$, and coalized all together in $\mathcal{N}$. Conventionally we set $v(\emptyset) = 0$ for the empty coalition.

A *cooperative* (or *coalitional*) *game* is defined by the pair $(\mathcal{N}, v)$ where the goal of each player is to optimize its value, accepting or not to take part in coalitions, including the grand coalition. The main question of coalitional games is the *fair allocation* of the total value $v(\mathcal{N})$ between all players, which is formalized by the concept of *value*, i.e. a function $\phi: \mathcal{G} \longrightarrow \mathbb{R}^{\mathrm{n}}$, where $\mathcal{G}$ is the set of all games with $n$ players, mapping each game $(\mathcal{N}, v)$ into a unique *allocation* $\mathcal{K} = \{K_1, \dots, K_n\}$ such that

$$\phi(\mathcal{N}, v) \longrightarrow \begin{bmatrix} \phi_1(\mathcal{N}, v) \\ \vdots \\ \phi_{\mathrm{n}}(\mathcal{N}, v) \end{bmatrix} = \begin{bmatrix} K_1 \\ \vdots \\ K_n \end{bmatrix}, \qquad K = \sum_{i \in \mathcal{N}} K_i = v(\mathcal{N}). \tag{1}$$

Notice that, in general,

$$v(\{i\}) \neq K_i, \qquad \sum_{i \in \mathcal{S} \subseteq \mathcal{N}} v(\{i\}) \neq v(\mathcal{S}) \neq \sum_{i \in \mathcal{S} \subseteq \mathcal{N}} K_i\,, \qquad \sum_{i \in \mathcal{N}} v(\{i\}) \neq v(\mathcal{N}) = \sum_{i \in \mathcal{N}} K_i = K. \tag{2}$$

If the characteristic function is subadditive and interpreted as a cost function, we have

$$v(\mathcal{S}_1 \cup \mathcal{S}_2) \leq v(\mathcal{S}_1) + v(\mathcal{S}_2), \tag{3}$$

leading to

$$K_i \leq v(\{i\}), \qquad \sum_{i \in \mathcal{S}} K_i \leq v(\mathcal{S}) \leq \sum_{i \in \mathcal{S}} v(\{i\})\,, \qquad \sum_{i \in \mathcal{N}} K_i = v(\mathcal{N}) \leq \sum_{i \in \mathcal{N}} v(\{i\}). \tag{4}$$

In this case players have an incentive to form coalitions $\mathcal{S}$, since this would reduce their total individual costs $\sum_{i \in \mathcal{S}} v(\{i\})$ to the total coalition cost $v(\mathcal{S})$. They only need to find a strategy to allocate the total cost $v(\mathcal{S})$ between members of $\mathcal{S}$, minimizing their individual costs $K_i$, with the obvious constrain $\sum_{i \in \mathcal{S}} K_i \leq v(\mathcal{S})$, otherwise every player in $\mathcal{S}$ would leave the coalition and find another better coalition.

## 2.2. Risk Allocation Theory

The game theory concepts summarized in the previous Sec. 2.1 may be adapted to a variety of different situations, provided that a translation, i.e. a one-to-one association is available between the two contexts.

In the context of risk allocation, one may identify the players with single *risk units*, i.e. risk factors (interest rates, equities, commodities, etc.), or single trades, or portfolios of trades, or business units. Coalitions are associated to groups of risk units. The grand coalition is associated to the total risk unit.

In order to make more explicit the uncertainty, i.e. the risk, associated to risk units, we assign to each player a *bounded random variable* $X_i: L^\infty(\Omega, \mathcal{F}, \mathbb{P}) \to \mathbb{R}$, where $L^\infty(\Omega, \mathcal{F}, \mathbb{P})$ is the space of bounded random variables, $(\Omega, \mathcal{F}, \mathbb{P})$ is a probability space, $\Omega$ is the set of all possible outcomes, $\mathcal{F}$ is the filtration including all subsets of $\Omega$ and $\mathbb{P}$ is a probability measure. We denote by $X_i$ the random variable representing i-th player with some probability distribution. For example, it can be interpreted as the value or the profit and loss of the i-th risk unit within a given time horizon *T*. In what follows we drop the time variable. The stochastic nature of the random variables $\mathcal{X} = \{X_1, \dots, X_n\}$, represents the risk associated to risk units. Therefore, the composite random variables

$$X_{\mathcal{S}} = \sum_{i \in \mathcal{S} \subseteq \mathcal{N}} X_i, \qquad X_N = \sum_{i \in \mathcal{N}} X_i = X, \tag{5}$$

represents the values associated to coalition $\mathcal{S}$ and to grand coalition $\mathcal{N}$, respectively. For consistency, we associate the null distribution $X_\emptyset$ to the null coalition (see Eq. (6) below).

We then associate the game characteristic function to a *risk measure* $\rho: \mathcal{X} \to \mathbb{R}$, which represents the level of risk of coalitions, such that

$$\rho(X_i) = v(\{i\}), \qquad \rho(X_{\mathcal{S}}) = v(\mathcal{S}), \qquad \rho(X) = v(\mathcal{N}) = K, \qquad \rho(X_\emptyset) = v(\emptyset) = 0, \tag{6}$$

are the risk measures associated to player $i$, coalition $\mathcal{S}$, grand coalition $\mathcal{N}$, and null coalition $\mathcal{S} = \emptyset$ respectively. The risk measure may be interpreted as a potential loss within time horizon *T* with a given confidence level, or as the corresponding minimum level of capital (cash) to be added to the coalition's worth or P&L to make them acceptable.

Finally, we define *risk allocation problem* the pair $(\mathcal{N}, \rho)$, i.e. the problem of allocating the amounts of risk $\{\rho(X_1), \dots, \rho(X_n)\}$ to players $\mathcal{N} = \{1, \dots, n\}$. We denote by $\mathcal{R}$ the set of all risk allocation problems with $n$ risk units, and define *risk allocation strategy* (a.k.a. *risk allocation principle*[2], *technique, approach,* or *scheme*) a function $\Pi: \mathcal{R} \to \mathbb{R}^n$ mapping each allocation problem $(\mathcal{N}, \rho)$ into a unique allocation $\mathcal{K} = \{K_1, \dots, K_n\}$,

$$\Pi(\mathcal{N}, \rho) = \begin{bmatrix} \Pi_1(\mathcal{N}, \rho) \\ \Pi_2(\mathcal{N}, \rho) \\ \vdots \\ \Pi_n(\mathcal{N}, \rho) \end{bmatrix} = \begin{bmatrix} K_1 \\ K_2 \\ \vdots \\ K_n \end{bmatrix}, \tag{7}$$

where $K_i$ stands for the allocated risk of the $i\text{-}th$ portfolio. A *fair risk allocation strategy* satisfies the *full allocation property*

$$\sum_{i \in \mathcal{N}} K_i = \rho(X). \tag{8}$$

Notice that, in general,

[2] In the literature "allocation principle" is frequently used. We prefer "allocation strategy" since the different allocation formulas are simply different ways to allocate risk, while "principle" refers to something more fundamental.

$$\rho(X_i) \neq K_i, \qquad \sum_{i\in\mathcal{S}} \rho(X_i) \neq \rho(X_\mathcal{S}) \neq \sum_{i\in\mathcal{S}} K_i\,, \qquad \sum_{i\in\mathcal{N}} \rho(X_i) \neq \rho(X) \tag{9}$$

The quantities

$$\rho(X_i) - K_i, \qquad \rho(X_\mathcal{S}) - \sum_{i\in\mathcal{S}} \rho(X_i)\,, \qquad \rho(X) - \sum_{i\in\mathcal{N}} \rho(X_i), \tag{10}$$

are the *diversification benefits* of the i-th player of coalition $\mathcal{S}$, and of grand coalition $\mathcal{N}$, respectively. We summarize in Table 1 the translation between game theory and risk allocation languages and notations.

Although in the following we will rely on the notation introduced above, in some practical applications the variables $\mathcal{X} = \{X_1, X_2, \dots, X_n\}$ may be represented by their actual (finite) empirical distributions, e.g. when they are estimated using Monte Carlo or historical scenarios. In these cases, each variable is redefined as follows

$$X_i \to \hat{X}_i = \{\hat{x}_{i,1}, \dots, \hat{x}_{i,m}\}, \tag{11}$$

where $\hat{X}_i$ is the estimated distribution at time *T* of $X_i$ with $m$ scenarios, $\hat{x}_{i,j}$ is the realization of variable $X_i$ in the scenario $j$,

$$\hat{X}_\mathcal{S} = \sum_{i\in\mathcal{S}\subseteq\mathcal{N}} \hat{X}_i\,, \qquad \hat{X}_\mathcal{N} = \sum_{i\in\mathcal{N}} \hat{X}_i = \hat{X}, \tag{12}$$

are the estimated distributions at time $T$ of $X_\mathcal{S}$ and $X_\mathcal{N}$, respectively, and

$$\hat{x}_{\mathcal{S},j} = \sum_{i\in\mathcal{S}\subseteq\mathcal{N}} \hat{x}_{i,j}\,, \qquad \hat{x}_{\mathcal{N},j} = \sum_{i\in\mathcal{N}} \hat{x}_{i,j}, \tag{13}$$

are the realizations of variable $X_\mathcal{S}$ and $X_\mathcal{N}$ in the scenario $j$.

| **Game Theory vs Risk Allocation languages and notations** | | |
|---|---|---|
| **Game Theory** | **Risk allocation** | **Examples** |
| Players<br>$\mathcal{P} = \{p_1, \dots, p_n\}$ | Risk units<br>$\mathcal{X} = \{X_1, \dots, X_n\}$ | • Single risk factor<br>• Single trade<br>• Single portfolio of trades<br>• Single business unit |
| Grand coalition<br>$\mathcal{N} = \{1, \dots, n\}$ | Total risk unit<br>$X_\mathcal{N} = \sum_{i\in\mathcal{N}} X_i$ | • All risk factors<br>• All trades<br>• All portfolios of trades<br>• Bank |
| Coalition<br>$\mathcal{S} = \{\emptyset, 1, \dots, s\} \subseteq \mathcal{N}$ | Group of risk units<br>$X_\mathcal{S} = \sum_{i\in\mathcal{S}\subseteq\mathcal{N}} X_i$ | • Group of risk factors<br>• Group of trades<br>• Group of portfolios of trades<br>• Group of business units |
| Characteristic function<br>$v: P(\mathcal{N}) \to \mathbb{R}$ | Risk measure<br>$\rho: \mathcal{X} \to \mathbb{R}$ | • Value at Risk<br>• Expected Shortfall<br>• Etc. see App. 8.1 |
| Cooperative game $(\mathcal{N}, v)$ | Risk allocation problem $(\mathcal{N}, \rho)$ | See above |
| Value $\phi(\mathcal{N}, v)$ | Risk allocation strategy $\Pi(\mathcal{N}, \rho)$ | • Marginal<br>• Shapley<br>• Etc. see Sec. 2.3 |
| Allocation $\mathcal{K} = \{K_1, \ldots, K_n\}$ | Allocation $\mathcal{K} = \{K_1, \ldots, K_n\}$ | -- |
| Fair game $\sum_{i \in \mathcal{N}} K_i = v(\mathcal{N})$ | Fair allocation strategy $\sum_{i \in \mathcal{N}} K_i = \rho(X)$ | -- |

***Table 1**: translation between game theory and risk allocation languages and notations.*

## 2.3. Allocation Strategies

Given a risk allocation problem $(\mathcal{N}, \rho)$, there are many different possible risk allocation strategies $\Pi(\mathcal{N}, \rho)$, whose properties depend on the properties of the risk measure $\rho$. We list below the most important examples. We report in App. 8.6 a comparison table which considers the following properties: full allocation, computational effort, interactions among risk units, potential negative values, theoretical or numerical issues.

### 2.3.1. Standalone

It consists in allocating to each risk unit $X_i$ an amount of risk excluding all the other contributors, namely

$$K_i^{\text{Sta}} = \rho(X_i). \tag{14}$$

Strictly speaking, this is not an allocation strategy, since it does not fulfill the full allocation property.

### 2.3.2. Proportional

It consists in allocating to each risk unit $X_i$ an amount of risk proportional to its own risk, namely

$$K_i^{\text{Pro}} = w_{\mathcal{N}}^{Pro} \rho(X_i), \qquad w_{\mathcal{N}}^{\text{Pro}} = \frac{\rho(X_{\mathcal{N}})}{\sum_{j \in \mathcal{N}} \rho(X_j)}. \tag{15}$$

The full allocation property is granted by construction through the normalization factor $w_{\mathcal{N}}^{\text{Pro}}$. Such strategy does not consider the dependency structure between the different risk units. However, it is frequently adopted because of its simplicity and ease of implementation. We will use it as a benchmark throughout the paper.

### 2.3.3. Marginal

It consists in allocating to each risk unit $X_i$ an amount of risk proportional to its marginal effect on the total risk unit. In other words, it is the impact on the total risk measure due to the exclusion of the i-th risk unit, namely

$$K_i^{\text{Mar}} = w_{\mathcal{N}}^{\text{Mar}} [\rho(X_{\mathcal{N}}) - \rho(X_{\mathcal{N}} - X_i)], \qquad w_{\mathcal{N}}^{\text{Mar}} = \frac{\rho(X_{\mathcal{N}})}{\sum_{j \in \mathcal{N}} [\rho(X_{\mathcal{N}}) - \rho(X_{\mathcal{N}} - X_j)]}. \tag{16}$$

The full allocation property is granted by construction through the normalization factor $w_{\mathcal{N}}^{\text{Mar}}$. Such method displays two main limitations: it does not consider the effect of the i-th risk unit on the other coalitions $\mathcal{S} \subseteq \mathcal{N}$, and it may lead to numerical instabilities if the marginal effect of the risk unit is small, i.e. when $\rho(X_{\mathcal{N}}) \simeq \rho\big(X_{\mathcal{N}} - X_j\big)$ and the denominator is small.

### 2.3.4. Shapley

It consists in allocating to each risk unit $X_i$ an amount of risk proportional to its *average* marginal effect[3] on all possible coalitions $\mathcal{S}$ which *include* the risk unit itself, namely

$$K_i^{\text{Sha}} = \sum_{\mathcal{S}\subseteq\mathcal{N}, i\in\mathcal{S}} w_{n,s}^{\text{Sha}}\big[\rho(X_{\mathcal{S}}) - \rho\big(X_{\mathcal{S}\setminus\{i\}}\big)\big], \qquad w_{n,s}^{\text{Sha}} = \frac{1}{n}\binom{n-1}{s-1}^{-1} = \frac{(s-1)!\,(n-s)!}{n!}, \tag{17}$$

where $s = |\mathcal{S}|$. Equivalently, one may take the average of marginal effects with respect to all possible coalitions $\mathcal{S}$ which *exclude* the risk unit itself, namely

$$K_i^{\text{Sha}} = \sum_{\mathcal{S}\subseteq\mathcal{N}\setminus\{i\}} w'^{\text{Sha}}_{n,s}\big[\rho\big(X_{\mathcal{S}\cup\{i\}}\big) - \rho(X_{\mathcal{S}})\big], \qquad w'^{\text{Sha}}_{n,s} = \frac{1}{n}\binom{n-1}{s}^{-1} = \frac{s!\,(n-s-1)!}{n!}. \tag{18}$$

Since $X_{\mathcal{S}\cup\{i\}} = X_{\mathcal{S}} + X_i$ and $X_{\mathcal{S}\setminus\{i\}} = X_{\mathcal{S}} - X_i$, we may also write

$$K_i^{\text{Sha}} = \sum_{\mathcal{S}\subseteq\mathcal{N}, i\in\mathcal{S}} w_{n,s}^{\text{Sha}}[\rho(X_{\mathcal{S}}) - \rho(X_{\mathcal{S}} - X_i)], \tag{19}$$

$$K_i^{\text{Sha}} = \sum_{\mathcal{S}\subseteq\mathcal{N}\setminus\{i\}} w'^{\text{Sha}}_{n,s}[\rho(X_{\mathcal{S}} + X_i) - \rho(X_{\mathcal{S}})]. \tag{20}$$

We point out that the sum in Eq. (19) excludes the empty coalition $\mathcal{S} = \emptyset, s = 0$ and includes the grand coalition $\mathcal{S} = \mathcal{N}, s = n$, while Eq. (20) includes $\mathcal{S} = \emptyset$ and excludes $\mathcal{S} = \mathcal{N}$.

It is evident that eq. (19) is a generalisation of the marginal allocation in eq. (16), also avoiding its potential unstable normalization factor. Contrary to Proportional and Marginal allocation strategies, Shapley considers the full interaction structure between all the risk units. In fact, we notice that the sum in Eq. (20) is performed for each of the $2^{n-1}$ coalitions $\mathcal{S} \subseteq \mathcal{N}\setminus\{i\}$, namely

$$\mathcal{S} = \emptyset, \{1\}, \{2\}, \dots, \{n\}, \{1,2\}, \dots, \{1, \dots, \{i\}, \dots, n\} \dots, \mathcal{N}\setminus\{i\}, \tag{21}$$

and its first and last terms $\mathcal{S} = \emptyset$ and $\mathcal{S} = \mathcal{N}\setminus\{i\}$ are given by

$$K_i^{\text{Sha}} = \frac{1}{n}\rho(X_i) + \frac{1}{n}[\rho(X_{\mathcal{N}}) - \rho(X_{\mathcal{N}} - X_i)] + \cdots \tag{22}$$

in which we recognize the Proportional and the Marginal allocations with weights $w'^{\text{Sha}}_{n,s=0} = w'^{\text{Sha}}_{n,s=n-1} = 1/n$. These two terms are the typically the largest contributions to the sum since their weights are the largest ones. The same result is obtained looking at Eq. (19). We show this effect in Figure 1.

The main drawback of Shapley allocation is the computational effort required by Eq. (19) or (20): for each risk unit $X_i$ the average over coalitions involves $2^{n-1}$ elements, and for each element the risk measure $\rho$ is computed twice. Since the double calculation of the risk measure can be avoided reusing previous calculations (see Sec. 3.1), the computational effort for the whole set of $n$ risk units grow as $n \times 2^{n-1}$, which becomes unmanageable already for a relatively low value of $n$, depending on the effort required to compute the risk measure $\rho$. More details on the computational cost are given in Sec. 3.2.

[3] This is the reason why we discuss Shapley allocation just after Marginal allocation.

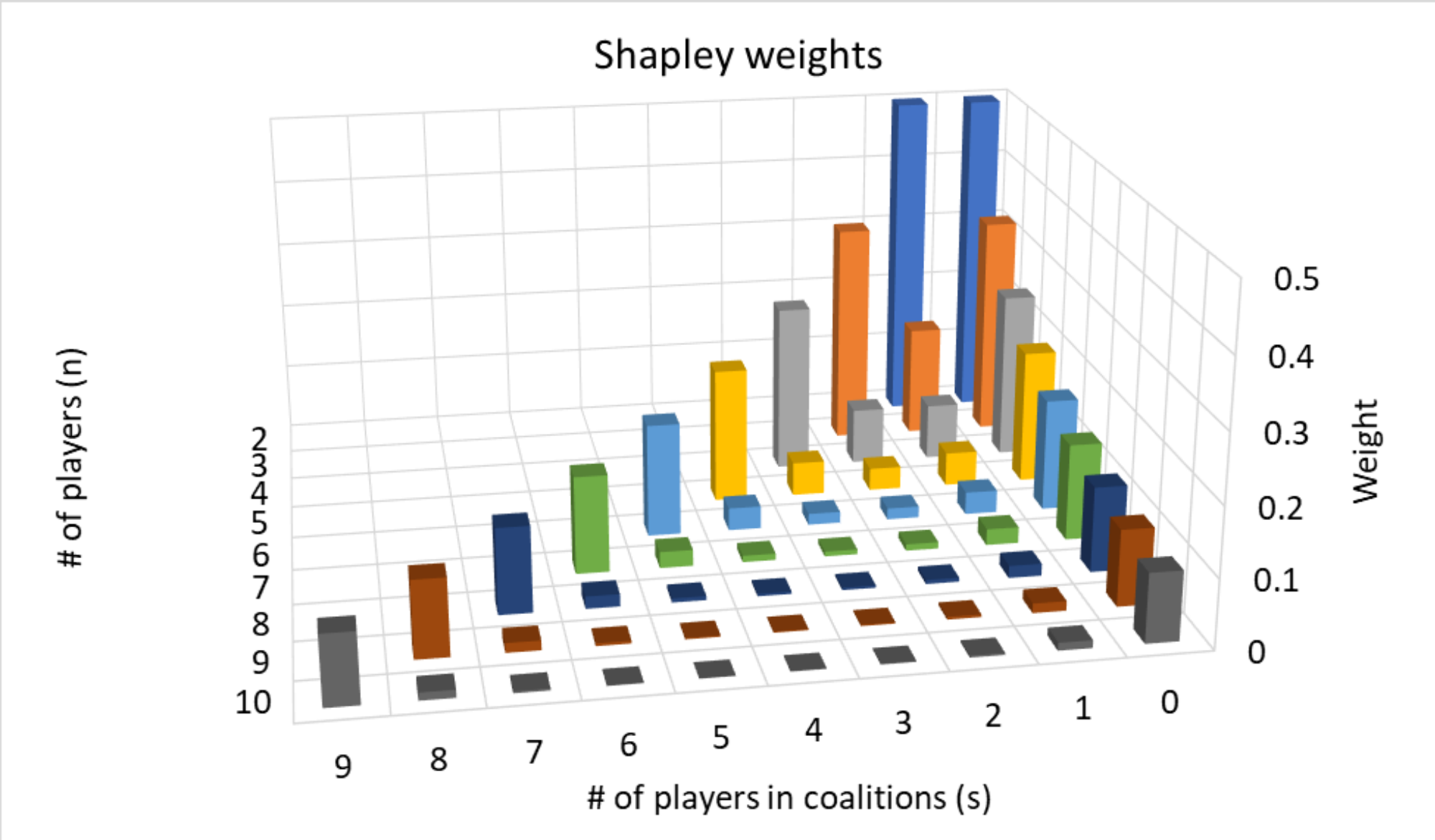

***Figure 1**: Shapley weights computed with Eq. (18). We notice that the largest weights are assigned, for any number of players $n$, to the empty coalition $\mathcal{S} = \emptyset, s = 0$, corresponding to the proportional allocation, and to the largest coalition $\mathcal{S} = \mathcal{N} \setminus \{i\}, s = n-1$, corresponding to the Marginal allocation. All the other terms contribute with smaller weights.*

### 2.3.5. Aumann-Shapley and Euler

**Aumann-Shapley** allocation strategy (Denault (2001)) is a generalization of Shapley allocation in case of fractional players, which may have a continuous level of presence in coalitions, such that

$$X(u) = \sum_{i=1}^{n} u_i X_i, \tag{23}$$

$$K_i^{\mathrm{AuSha}} = \int_0^1 \frac{\partial \rho[X(\lambda u)]}{\partial(\lambda u_i)} d\lambda \Bigg|_{u=1}, \tag{24}$$

where the weight variables $u = \{u_1, \dots, u_n\} \in [0,1]^n$ represent the fractional worth associated with each risk unit, or, in other words, the amount of money invested in the underlying assets. From Eq. (24) we recover Eq. (5) with $X = X(u = 1)$**.** We may look at eq. (24) as the infinitesimal version of Shapley allocation in eq. (20), i.e. a continuous sum of infinitesimal marginal contributions

$$\frac{\partial \rho[X(\tilde{u})]}{\partial \tilde{u}_i} := \lim_{\Delta \tilde{u}_i \to 0} \frac{\rho[X(\tilde{u} + \Delta \tilde{u}_i)] - \rho[X(\tilde{u})]}{\Delta \tilde{u}_i}, \tag{25}$$

where $\tilde{u} = \lambda u$, and $X(\tilde{u} + \Delta \tilde{u}_i) = \sum_{j=1}^{n} (\tilde{u}_j + \Delta \tilde{u}_i \delta_{ij}) X_j$.

**Euler** allocation strategy is a particular case of Aumann-Shapley allocation for first-order homogeneous and differentiable risk measures, such that $\rho[X(\lambda u)] = \lambda \rho[X(u)]$. In this case, as shown in Tasche (2008), the allocation formula reads

$$K_i^{\text{Eul}} = \int_0^1 \frac{\partial \rho[X(\lambda u)]}{\partial(\lambda u_i)} d\lambda \Bigg|_{u=1} = \int_0^1 \frac{\partial \rho[X(u)]}{\partial(\lambda u_i)} \lambda d\lambda \Bigg|_{u=1} = \frac{\partial \rho[X(u)]}{\partial u_i}\Bigg|_{u=1}. \tag{26}$$

The full allocation property is granted by the Euler theorem on homogeneous functions (Tasche (2008)),

$$\sum_{i=1}^{n} K_i^{\text{Eul}} = \sum_{i=1}^{n} u_i K_i^{\text{Eul}}\Bigg|_{u=1} = \sum_{i=1}^{n} u_i \frac{\partial \rho[X(u)]}{\partial u_i}\Bigg|_{u=1} = \rho(X). \tag{27}$$

Since the direct implementation of Eq. (26) may be cumbersome, Tasche (2008) provides specific formulas for some popular risk measures, described below.

**Euler Standard Deviation**
We have

$$\rho(X) = \text{Std(X)}, \qquad K_i^{\text{Eul,Std}} = \frac{\text{cov}(X_i, X)}{\sqrt{\text{var}(X)}}, \tag{28}$$

where $\text{var}, \text{cov}$ are the variance and covariance functions.

**Euler Expected Shortfall**
We have, for confidence level[4] $\alpha$,

$$\rho(X) = \text{ES}_\alpha(\text{X}), \qquad K_i^{\text{Eul,ES}} = -(1-\alpha)\mathbb{E}\left[X_i 1_{\{X \le -\text{VaR}_\alpha(X)\}}\right], \tag{29}$$

where $\text{VaR}_\alpha$ is the Value at Risk at confidence level $\alpha$, $1_{\{\cdot\}}$ is the indicator function and $\mathbb{E}[\cdot]$ is the expected value operator.

**Euler VaR**
We have, for confidence level $\alpha$,

$$\rho(X) = \text{VaR}_\alpha(\text{X}), \qquad K_i^{\text{Eul,VaR}} = -\mathbb{E}[X_i | X = -\text{VaR}_\alpha(X)]. \tag{30}$$

As pointed out in Tasche (2008), the direct computation of the expectation in Eq. (30) may be challenging. For this reason, the following **first order approximation** is available, based on a linear approximation of the expected value

$$K_i^{\text{Eul,VaR,1st}} = \mathbb{E}[X_i] + \frac{\text{cov}(X_i, X)}{\text{var}(X)}\left[\text{VaR}_\alpha(X) - \mathbb{E}[X]\right]. \tag{31}$$

[4] Throughout this paper, Value at Risk is intended as a percentile (left tail) of the profit and loss distribution.

Being Eq. (31) derived as an approximation, the full allocation property is no longer granted. To improve the quality of the approximation, Tasche and Tibiletti (2001) propose a **second order approximation**,

$$K_i^{\text{Eul,VaR,2nd}} = \mathbb{E}[X_i] + a_i\big[\text{VaR}_\alpha(X) - \mathbb{E}[X]\big] + b_i\big[\text{VaR}_\alpha(X^2) - \mathbb{E}[X^2]\big], \tag{32}$$

where

$$a_i = \frac{\text{var}(X^2)\text{cov}(X_i, X) - \text{cov}(X, X^2)\text{cov}(X_i, X^2)}{\text{var}(X)\text{var}(X^2) - \text{cov}(X, X^2)^2}, \tag{33}$$

$$b_i = \frac{\text{var}(X)\text{cov}(X_i, X^2) - \text{cov}(X, X^2)\text{cov}(X_i, X)}{\text{var}(X)\text{var}(X^2) - \text{cov}(X, X^2)^2}. \tag{34}$$

As discussed in Sec. 2.2, for many practical applications the variables $\{X_1, \dots, X_n\}$ are represented by their actual (finite) empirical distributions defined in Eq. (11), $\hat{X}_i = \{\hat{x}_{i,1}, \dots, \hat{x}_{i,m}\}$. According to Tasche (2008), in these situations it is possible to obtain consistent estimators of the Euler allocation simply substituting $X$ with $\hat{X}$ for Standard Deviation (Eq. (29)) and Expected Shortfall (Eq. (30)). In contrast, for Value at Risk this naïve approach does not work. A possible workaround is based on the **Nadaraya-Watson Kernel** as shown below

$$K_i^{\text{Eul,VaR,Ker}} = -\frac{\sum_{k=1}^{m} \hat{x}_{i,k}\,\kappa\left(\frac{-\text{VaR}_\alpha(\hat{X}+b) - \hat{x}_k}{b}\right)}{\sum_{k=1}^{m} \kappa\left(\frac{-\text{VaR}_\alpha(\hat{X}+b) - \hat{x}_k}{b}\right)}, \tag{35}$$

where $\kappa$ is the (Gaussian) kernel and $b$ is estimated as follows

$$b = 0.9\,\min(\sigma, R/1.34)\,n^{-1/5}, \tag{36}$$

where $\sigma$ and $R$ are the standard deviation and the interquartile range of the distribution $\hat{x}_1, \dots, \hat{x}_m$, respectively.

## 2.4. Negative Shapley Allocation

Shapley allocation may lead to negative values when, adding a risk unit to a coalition $\mathcal{S}$ reduces the risk of the coalition, i.e. $\rho(X_\mathcal{S} + X_i) < \rho(X_\mathcal{S})$. In practice, this effect occurs either for risk units hedging many other risk units, or when the risk unit is negligible in many coalitions. While negative Shapley allocations are theoretically meaningful, they can be problematic in case of allocation of positive risk measures. To solve this issue Denault (2001) invokes a pragmatic “crossed finger” approach, assuming that in real-life the issue may be ignored since quite unlikely. Liu (2020) proposes two possible modifications of the original Shapley formula forcing non-negative values,

$$K_i^{\text{Sha}} = \sum_{\mathcal{S}\subseteq\mathcal{N}, i\in\mathcal{S}} w_i^{\text{Sha}} |\rho(X_\mathcal{S}) - \rho(X_\mathcal{S} - X_i)|, \tag{37}$$

$$K_i^{\text{Sha}} = \sum_{\mathcal{S}\subseteq\mathcal{N}, i\in\mathcal{S}} w_i^{\text{Sha}} \max[\rho(X_\mathcal{S}) - \rho(X_\mathcal{S} - X_i); 0]. \tag{38}$$

Unfortunately, both the approaches above are not backed by any theoretical foundation and break the full allocation property. Instead, we propose below two simple alternative approaches to manage negative Shapley allocations while preserving the full allocation property. The set negative Shapley allocation set is denoted hereafter by $\mathcal{N}^- = \{i \in \mathcal{N} \text{ s.t.} K_i^{\text{Sha}} < 0\}$.

### 2.4.1. Shapley Maximum Proportional

We set to zero any negative allocation,

$$K_i^{\text{Sha}+} = \max\left[K_i^{\text{Sha}}; 0\right], \tag{39}$$

and redistribute the resulting non-negative allocations proportionally to all risk units,

$$K_i^{\text{ShaMax}} = \frac{\rho(X)}{\sum_{i\in\mathcal{N}} K_i^{\text{Sha}+}} K_i^{Sha+} = \frac{\rho(X)}{\rho(X) - \sum_{i\in\mathcal{N}^-} K_i^{\text{Sha}}} K_i^{\text{Sha}+}. \tag{40}$$

From Eq. (40) we have that

$$\begin{cases} K_i^{\text{ShaMax}} \leq K_i^{\text{Sha}} & \text{for } i \notin \mathcal{N}^-, \\ K_i^{\text{ShaMax}} = 0 & \text{for } i \in \mathcal{N}^-. \end{cases} \tag{41}$$

The resulting allocation in Eq. (40) above, which we call *Shapley Maximum Proportional*, respects the full allocation property,

$$\sum_{i\in\mathcal{N}} K_i^{\text{ShaMax}} = \frac{\rho(X)}{\rho(X) - \sum_{i\in\mathcal{N}^-} K_i^{\text{Sha}}} \sum_{i\in\mathcal{N}} K_i^{\text{Sha}+} = \frac{\rho(X)}{\rho(X) - \sum_{i\in\mathcal{N}^-} K_i^{\text{Sha}}} \left(\rho(X) - \sum_{i\in\mathcal{N}^-} K_i^{\text{Sha}}\right) = \rho(X). \tag{42}$$

### 2.4.2. Shapley Absolute Proportional

We take the absolute value of any negative allocation,

$$K_i^{|\text{Sha}|} = \left|K_i^{\text{Sha}}\right|, \tag{43}$$

and redistribute the resulting positive allocations proportionally to all risk units

$$K_i^{\text{ShaAbs}} = \frac{\rho(X)}{\sum_{i\in\mathcal{N}} K_i^{|\text{Sha}|}} K_i^{|\text{Sha}|} = \frac{\rho(X)}{\rho(X) - 2\sum_{i\in\mathcal{N}^-} K_i^{\text{Sha}}} K_i^{|\text{Sha}|} \tag{44}$$

From Eq. (44) we have that

$$\begin{cases} K_i^{\text{ShaAbs}} \leq K_i^{\text{Sha}} & \text{for } i \notin \mathcal{N}^- \\ K_i^{\text{ShaAbs}} = |K_i^{\text{Sha}}| & \text{for } i \in \mathcal{N}^-. \end{cases} \tag{45}$$

The resulting allocation in Eq. (44) above, which we call *Shapley Absolute Proportional*, respects the full allocation property, since

$$\sum_{i\in\mathcal{N}} K_i^{\text{ShaAbs}} = \frac{\rho(X)}{\sum_{i\in\mathcal{N}} K_i^{|\text{Sha}|}} \sum_{i\in\mathcal{N}} K_i^{|\text{Sha}|} = \rho(X). \tag{46}$$

In Sec. 4.4 we show some numerical examples of negative Shapley allocation.

## 2.5. Multi-Level Allocation Strategies

The allocation strategies discussed in Sec. 2 consider $n$ distinct risk units organised in a single hierarchical level, referred to as the *Single-Level allocation strategy*. Clearly, financial institutions exhibit complex organizational structures spanning multiple hierarchical levels, where the risk units of the lower hierarchical level belong to the corresponding parent units of the upper level, culminating with the topmost hierarchical level of the firm. In general, the number of distinct risk units decreases (increases) moving up (down) in the hierarchy, and risk units at each layer, except the lowest ones, encompass a variable number of risk units from the level below. We refer to the allocation problem applied to such composite situation as *Multi-Level allocation strategy*. Since the Multi-Level allocation problem can be viewed as multiple Single-Level allocations applied to each pair of consecutive layers of the firm's organizational structure, the following formalization considers a generic upper layer $U$ and a lower layer $L$ such that $U = L + 1$. The generalization to all levels is straightforward.

Following the notation introduced in Sec. 2.2, we consider the variables $\mathcal{X}^U = \{ X_1^U, \dots, X_{n_U}^U\}$ and $\mathcal{X}^L = \{ X_1^L, \dots, X_{n_L}^L\}$ made by $n_U$ and $n_L$ distinct risk units, with $n_U < n_L$. Each risk unit of $\mathcal{X}^U$ is composed by a collection of risk units of $\mathcal{X}^L$ such that

$$X_i^U = \sum_{j \in \mathcal{I}(X_i^U)} X_j^L \text{ for } i = 1, \dots, n_U \tag{47}$$

where $\mathcal{I}(X_i^U)$ denotes the set of indices of child risk units of $\mathcal{X}^L$ corresponding to the parent risk unit $X_i^U$. To ensure the consistency between the two hierarchy layers, we impose that the sets $\mathcal{I}(X_i^U)$ are a partition of $\{1, \dots, n_U\}$, namely

$$\bigcap_{i=1}^{n_U} \mathcal{I}(X_i^U) = \emptyset, \qquad \bigcup_{i=1}^{n_U} \mathcal{I}(X_i^U) = \{1, \dots, n_U\}. \tag{48}$$

We point out that from Eq. (5) and Eq. (48) we have that

$$\sum_{i=1}^{n_L} X_i^L = \sum_{i=1}^{n_U} X_i^U = X. \tag{49}$$

The application of the Single-Level allocation strategy to levels $U$ and $L$ under the risk measure $\rho$ with a generic allocation strategy $\Pi$ leads to the (independent) allocations $\mathcal{K}^L = \{K_1^L, \dots, K_{n_L}^L\}$ and $\mathcal{K}^U = \{K_1^U, \dots, K_{n_U}^U\}$. From Eq. (49) and the assumption that $\Pi$ is a fair risk allocation strategy, we have that both allocations satisfy the Single-Level full allocation property, namely

$$\sum_{i=1}^{n_L} K_i^L = \sum_{i=1}^{n_U} K_i^U = \rho(X) = K \tag{50}$$

where $\rho(X)$ is the overall risk measure of the firm introduced in (6). However, in general, the full allocation property is not guaranteed for each parent $X_i^U$, namely

$$\sum_{j \in \mathcal{I}(X_i^U)} K_j^L \neq K_i^U. \tag{51}$$

Eq. (51) shows that the allocation of a parent portfolio is not the simple sum of the allocations of the child portfolios. The lack of the sub-additivity expressed in Eq. (51) can be problematic depending on business constraints driven by specific business mandates or risk management protocols. Hence, we look for suitable adjusted allocations $\mathcal{K}^L$ or $\mathcal{K}^U$ that overcome Eq. (51), such that

$$\sum_{j \in \mathcal{I}(X_i^U)} K_j^L = K_i^U \ \forall i. \tag{52}$$

We refer to Eq. (52) as *Multi-Level full allocation property*. We provide below three possible approaches.

### 2.5.1. Proportional Top-Down (PTD) Approach

The *Proportional Top-Down* approach (PTD) is based on the adjusted child allocations $\mathcal{K}^{L,\mathrm{PTD}}$ that take into account the importance of each child $K_j^L$ for $j \in \mathcal{I}(X_i^U)$ to each corresponding parent $X_i^U$ in a proportional fashion, namely

$$K_j^{L,\mathrm{PTD}} = \frac{K_j^L}{\sum_{j \in \mathcal{I}(X_i^U)} K_j^L} K_i^U. \tag{53}$$

Clearly, we have that

$$\sum_{j \in \mathcal{I}(X_i^U)} K_j^{L,\mathrm{PTD}} = K_i^U, \qquad \sum_{j=1}^{n_L} K_j^{L,\mathrm{PTD}} = K. \tag{54}$$

where the second equation reads from (48) and (54) as follows

$$\sum_{j=1}^{n_L} K_j^{L,\mathrm{PTD}} = \sum_{i=1}^{n_U} \sum_{j \in \mathcal{I}(X_i^U)} K_j^{L,\mathrm{PTD}} = \sum_{i=1}^{n_U} \left[ \frac{\sum_{j \in \mathcal{I}(X_i^U)} K_j^L}{\sum_{j \in \mathcal{I}(X_i^U)} K_j^L} \right] K_i^U = \sum_{i=1}^{n_U} K_i^U = K. \tag{55}$$

We observe that the PTD approach requires the recursive downward application of the allocation strategy $\Pi$ to each hierarchical level, and that the sub-additivity across hierarchical levels is guaranteed by a simple reproportioning, irrespective of the allocation strategy $\Pi$ adopted.

### 2.5.2. Constrained Top-Down (CTD) Approach

The *Constrained Top-Down* approach (CTD), proposed by Lin et al. (2016) and applied in Li, Xing (2019) in the context of FRTB, is based on the calculation of level $L$ allocations $K_j^{L,\mathrm{CTD}}$ for $j \in \mathcal{I}(X_i^U)$ adopting the same allocation strategy $\Pi$ adopted to compute $K_i^U$, but constraining the scope of the computation to $X_i^U$ and its sub risk units. In such a way Eq. (55)(57) is satisfied by construction.

We observe that the CTD approach requires the application of the allocation strategy $\Pi$ to every hierarchical level without the introduction of the proportional technique of PTD approach. However, the restriction of the scope of the problem prevents capturing the relationships among risk units occurring in level $L$ but belonging to different parent risk units, that could be a limitation if no specific constraints about the cross desk operativity exist.

When applied to Shapley allocation strategy, CTD approach restricts the permutations of different risk units to each parent business unit. Consequently, it significantly reduces computational effort, as the sum of calculations required for each parent business unit is much lower than the total number of calculations needed if all permutations of portfolios at the same hierarchical level were considered. Finally, we note that if $\Pi$ is the Proportional allocation strategy (Sec. 2.3.2), CTD and PTD coincide.

### 2.5.3. Bottom-Up Approach

The *Bottom-Up* approach (BU) is based on the update of the allocations of $\mathcal{K}^U$, whose adjusted version $\mathcal{K}^{U,\mathrm{BU}}$ is computed summing up the allocations of the child level $L$ for $i = 1, \dots, n_U$ belonging to the risk unit $X_i^U$, namely

$$K_i^{U,\mathrm{BU}} = \sum_{j \in \mathcal{I}(X_i^U)} K_j^L. \tag{56}$$

Clearly, the definition of Eq. (56) solves the issue of Eq. (51); moreover, analogously to Eq. (55), we have that

$$\sum_{i=1}^{n_U} K_i^{U,\mathrm{BU}} = K. \tag{57}$$

We observe that the BU approach, in contrast with the PTD one, only requires the application of the allocation strategy $\Pi$ to the lowest level, applying Eq. (56) recursively to all the upper ones, losing the information carried out by the allocation strategy $\Pi$.

## 2.6. Hybrid Allocation Strategies

A composite risk measure may include different pieces which capture different risk classes. This is the case, for example, of market risk capital charge both under Basel 2.5 and FRTB regulations, as discussed in Apps. 8.4 and 8.5, respectively. In both cases the total risk measure includes a default risk component (IRC and DRC, respectively) based on computationally expensive Monte Carlo simulations. In these cases, the risk allocation task results to be prohibitive, and one must resort to some practical workaround.

The simplest approach is to use a *hybrid allocation strategy* based on Shapley allocation for the tractable part of the capital charge, and the Proportional allocation strategy for the intractable part. For example, in the case of Basel 2.5 capital charge (see App. 8.4), one may compute Shapley allocations $K_i^{\mathrm{Sha}}$ for VaR+sVaR in Eq. (77) and Proportional allocation for IRC, namely

$$K_i = K_i^{\mathrm{Sha\ VaR+sVaR}} + \frac{\mathrm{IRC}(X)}{\sum_{i=1}^{n} \mathrm{IRC}(X_i)} \mathrm{IRC}(X_i), \tag{58}$$

which requires only the calculation of $n + 1$ allocations. The same hybrid allocation strategy may be used for FRTB IMA capital charge in Eq. (79).

# 3. Numerical Methods

In this section we delve into various numerical aspects involved in the computation of the allocation strategies described in Sec. 2.3, with a specific focus on Shapley allocation. In particular, Sec. 3.1 provides a detailed description of the calculation of Shapley Monte Carlo estimate. Sec. 3.2 investigates the computational cost associated with the application of Shapley allocation, while Sec. 3.3 addresses how to calculate marginal contributions in the context of empirical distributions, e.g. in case of historical scenarios of the historical VaR.

## 3.1. Shapley Monte Carlo

When the computation of the analytical Shapley allocation discussed in Sec. 2.3.4 is practically unfeasible because of the large number of possible coalitions, it is possible to resort to a Monte Carlo approach, originally introduced in Mann and Shapley (1960). Following e.g. Hagan et al. (2021), Rozemberczki et al. (2022) and Mitchell et al. (2022), we rewrite the Shapley value of Eq. (20) in a more convenient way. We denote with $\pi \in P(\mathcal{N})$ a generic element of the power set $P(\mathcal{N})$, and with $\pi(i) = j$ the position $j$ (rank) of the $i$-th risk unit in the permutation $\pi$. For example, let $n = 3$, $\mathcal{N} = \{1,2,3\}$, $P(\mathcal{N}) = \{\{1,2,3\},\{1,3,2\},\{2,1,3\},\{2,3,1\},\{3,1,2\},\{3,2,1\}\}$ and consider the permutation $\pi = \{2,1,3\}$: we have $\pi(1) = 2$, $\pi(2) = 1$, $\pi(3) = 3$. We also introduce the inverse $\pi^{-1}(j) = i$, which, for a given position $j$ of the $i$-th risk unit in the permutation $\pi$, gives the corresponding risk unit $i$. Hence, we define the *predecessor set* $P_i^\pi$ of the $i$-th risk unit relative to the permutation $\pi$ as the set of risk units ranked lower than $i$ in $\pi$, namely

$$P_i^\pi = \{j \in \mathcal{N} \mid \pi(j) < \pi(i)\}. \tag{59}$$

For example, if $n = 3$ and $\pi = \{2,1,3\}$ we have $P_1^\pi = \{2\}, P_2^\pi = \emptyset, P_3^\pi = \{2,1\}$. It is now possible to write equivalently the Shapley allocation of Eq. (20) as the average marginal contributions of $i$-th risk unit to the risk of the predecessor set with respect to any possible permutation,

$$K_i^{\text{Sha}} = \frac{1}{n!} \sum_{\pi \in P(\mathcal{N})} \left[\rho\left(X_{P_i^\pi} + X_i\right) - \rho\left(X_{P_i^\pi}\right)\right]. \tag{60}$$

The Shapley formula in Eq. (60) above is more symmetric than Eq. (20), but it involves $n!$ terms, much more than $2^{n-1}$. Now we may define the Monte Carlo estimate of Eq. (60) as follows

$$K_i^{\text{Sha,MC}} = \frac{1}{N_{MC}} \sum_{\pi \in \mathrm{P}_{MC}} \left[\rho\left(X_{P_i^\pi} + X_i\right) - \rho\left(X_{P_i^\pi}\right)\right], \tag{61}$$

where $\mathrm{P}_{MC} \subset P(\mathcal{N})$ is a uniform sample of $N_{MC}$ permutations of size $n$. As discussed in Mitchell et al. (2022), $K_i^{\text{Sha,MC}}$ in Eq. (61) is an unbiased estimator that converges asymptotically to the Shapley value $K_i^{\text{Sha}}$ of Eq. (20) at a rate of $O(1/\sqrt{N_{\text{MC}}})$.

The computational effort required to calculate Eq. (61) above grows linearly with the number $n$ of risk units. A non-optimized algorithm would require $n \times N_{MC}$ permutation samples and $2 \times n \times N_{MC}$ calculations of the risk measure $\rho$ to converge at rate $O(1/\sqrt{N_{MC}})$. Actually, the calculation can be speeded up in a number of ways. First, a single sample of $\pi(N)$ can be used to compute all the Shapley values $K_i^{Sha}$ $\forall i$ in a single sweep. Second, non-necessary double calculations of the risk measure $\rho$ can be avoided. Furthermore, Mitchell et al. (2022) propose several techniques to improve the Monte Carlo convergence rate, i.e. to efficiently sample permutations in $P(\mathcal{N})$. In this work we adopt the antithetic sampling technique described by Ross (2005), considering the permutation sample $\pi$ and its reverse $\hat{\pi}$,

$$\pi = \{\pi(1), \pi(2), \dots, \pi(n-1), \pi(n)\} \longrightarrow \hat{\pi} = \{\pi(n), \pi(n-1), \dots, \pi(2), \pi(1)\}. \tag{62}$$

All these features are included in the Algorithm 1 below, which requires $N_{MC}/2$ permutation samples and $n \times N_{MC}/2$ calculations of the risk measure to converge at the same rate $O(1/\sqrt{N_{\text{MC}}})$.

**Algorithm 1:** Shapley Monte Carlo permutation sampling

1. Initialize[5] $[K_{1,0}, \dots, K_{n,0}] = [0, \dots, 0]$
2. Loop on MC scenarios $m = 1, \dots, N_{MC}$
   2.1. Initialize $\rho_{old} = \rho(\emptyset) = 0$
   2.2. Sample $\pi$ from $\Pi(\mathcal{N})$
   2.3. Compute $\hat{\pi}$
   2.4. Loop on risk units $i = 1, \dots, n$
      2.4.1. Set $\alpha = \pi^{-1}(i), \hat{\alpha} = \hat{\pi}^{-1}(i)$,
      2.4.2. Compute $\rho_{new} = \frac{1}{2}\left[\rho\left(X_{P_\alpha^\pi} + X_\alpha\right) + \rho\left(X_{P_{\hat{\alpha}}^\pi} + X_{\hat{\alpha}}\right)\right]$
      2.4.3. Compute $K_{\alpha,m} = K_{\alpha,m-1} + \rho_{new} - \rho_{old}$
      2.4.4. Compute $K_{\alpha,m}^2$
      2.4.5. Set $\rho_{old} = \rho_{new}$
3. Compute the MC averages $K_i = \frac{K_{i,N_{MC}}}{N_{MC}},\ i = 1, \dots, n$
4. Compute the MC standard errors $\epsilon_i = \sqrt{\frac{K_{i,N_{MC}}^2}{N_{MC}} - K_i^2}, i = 1, \dots, n$

We observe that in step 2.4.2 of Algorithm 1 we included the antithetic sampling, while in the loop 2.4 we update the whole Shapley value vector $K_{i,m}$ using a single permutation sampling $\pi$. Moreover, in step 2.4.5 we store the value of the risk measure $\rho_{old}$ corresponding to permutations $\pi^{-1}(i), \hat{\pi}^{-1}(i)$ and we reuse it for the computation of the following $K_{i+1,m}$ thus avoiding to double the number of risk measure computations.

More remarkably, Algorithm 1 above grants by construction the full allocation property: for each single permutation sample the sum of Shapley allocations is constant and equal to the overall measure, and the only quantities that change during the Monte Carlo simulation are the allocation proportions of the risk units. In formulas,

$$\rho(X) = \sum_{i=1}^{n} K_{i,m} = K \ \ \forall m = 1, \dots, N_{MC}, \tag{63}$$

where $K_{i,m}$ is the Shapley value for risk unit $i$ at path $m$. We provide the proof of this fact in App. 8.2, while extensive numerical tests on the Shapley Monte Carlo strategy are reported in Sec. 4.3.

## 3.2. Computational Cost

The computational cost of the allocation strategies discussed in Sec. 2.3 and 3.1 are summarized in the following Table 2.

| Allocation strategy | Number of computations of risk measure $\rho$ |
|---|---|
| Standalone | $n$ |
| Proportional | $n$ |
| Marginal | $n$ |
| Shapley | $n \times 2^{n-1}$ |
| Shapley Monte Carlo (antithetic) | $n \times N_{MC}/2$ |
| Euler | $n$ |

***Table 2**: computational cost of different allocation strategies in terms of computations of the risk measure $\rho$, as a function of number $n$ of risk units and number $N_{MC}$ of Monte Carlo samples. The number of computations associated to Shapley assumes a naive implementation without any optimization.*

[5] We initialize the MC Shapley estimates of the zeroth path to zero.

As discussed in Sec. 2.3.4 and 3.1, the computational cost of the Shapley allocation strategy for $n$ risk units, measured in terms of calculations of the risk measure $\rho$, grows exponentially as $n \times 2^{n-1}$ for Shapley analytical formulas and linearly as $n \times N_{MC}/2$ for Shapley Monte Carlo (which converges at rate $O(1/\sqrt{N_{\mathrm{MC}}})$ using $N_{MC}/2$ permutation samples). The actual computational cost depends on the effort required to compute the risk measure $\rho$. We show in Table 3 below an example of computational cost in a specific setting. As expected, ratios between adjacent columns tend to adapt, for each $m$, to the theoretical ratio $(n+1)2^{n+1}/n2^n = 2(n+1)/n$. Clearly, Shapley Monte Carlo becomes rapidly the only feasible choice, which, once adopted algorithm 1 in Sec. 3.1, scales as $n \times N_{MC}/2$.

| | n | | | | | | | | | | | | | | | | | | | | | | | |
|---|---|---|---|---|---|---|---|---|---|---|---|---|---|---|---|---|---|---|---|---|---|---|---|---|
| **m** | **4** | **5** | **6** | **7** | **8** | **9** | **10** | **11** | **12** | **13** | **14** | **15** | **16** | **17** | **18** | **19** | **20** | **21** | **22** | **23** | **24** | **25** | **26** | **27** |
| **250** | 0.0 | 0.0 | 0.0 | 0.0 | 0.1 | 0.2 | 0.3 | 0.8 | 1.7 | 3.6 | 7.8 | 16.7 | 35.7 | 75.8 | 2.7 | 5.6 | 11.9 | 25.0 | 52.3 | 1.8 | 3.8 | 7.9 | 16.5 | 1.4 |
| **500** | 0.0 | 0.0 | 0.0 | 0.0 | 0.1 | 0.2 | 0.5 | 1.1 | 2.3 | 5.0 | 10.8 | 23.1 | 0.8 | 1.7 | 3.7 | 7.8 | 16.4 | 34.5 | 1.2 | 2.5 | 5.3 | 11.0 | 22.8 | 2.0 |
| **1000** | 0.0 | 0.0 | 0.1 | 0.1 | 0.3 | 0.6 | 1.4 | 3.1 | 6.7 | 14.4 | 31.1 | 1.1 | 2.4 | 5.0 | 10.6 | 22.5 | 47.3 | 1.7 | 3.5 | 7.3 | 15.1 | 1.3 | 2.7 | 5.7 |
| **10000** | 0.1 | 0.2 | 0.4 | 0.9 | 2.1 | 4.7 | 10.4 | 22.9 | 50.0 | 1.8 | 3.9 | 8.3 | 17.8 | 37.8 | 1.3 | 2.8 | 5.9 | 12.4 | 1.1 | 2.3 | 4.7 | 9.9 | 20.5 | 42.7 |
| **100000** | 0.4 | 1.1 | 2.7 | 6.3 | 14.4 | 32.3 | 71.9 | 2.6 | 5.7 | 12.5 | 26.8 | 57.5 | 2.0 | 4.3 | 9.2 | 19.4 | 1.7 | 3.6 | 7.5 | 15.7 | 32.7 | 68.1 | 141.7 | 294.3 |
| | seconds | | | | | | | | minutes | | | | | | | hours | | | | | | days | | |

***Table 3**: computational cost (seconds: green, minutes: yellow, hours: orange, days: red) of analytical Shapley allocation for Value at Risk computed as described in Sec. 4.1 as a function of number $n$ of risk units and number $m$ of gaussian samples. Figures obtained with Matlab 2022 on Intel Core i5-8365U CPU @ 1.60GHz. The largest points were estimated as $n \times 2^n \times \Delta T_\rho(m)$, where $\Delta T_\rho(m)$ is the time required to compute $\rho$ for a given $m$.*

In Figure 2 we show the scaling laws of Shapley and Shapley Monte Carlo allocation strategies with the number $n$ of risk units. We observe that Shapley Monte Carlo becomes preferable already for relatively small $n \simeq 10 - 14$ depending on the number of Monte Carlo scenarios.

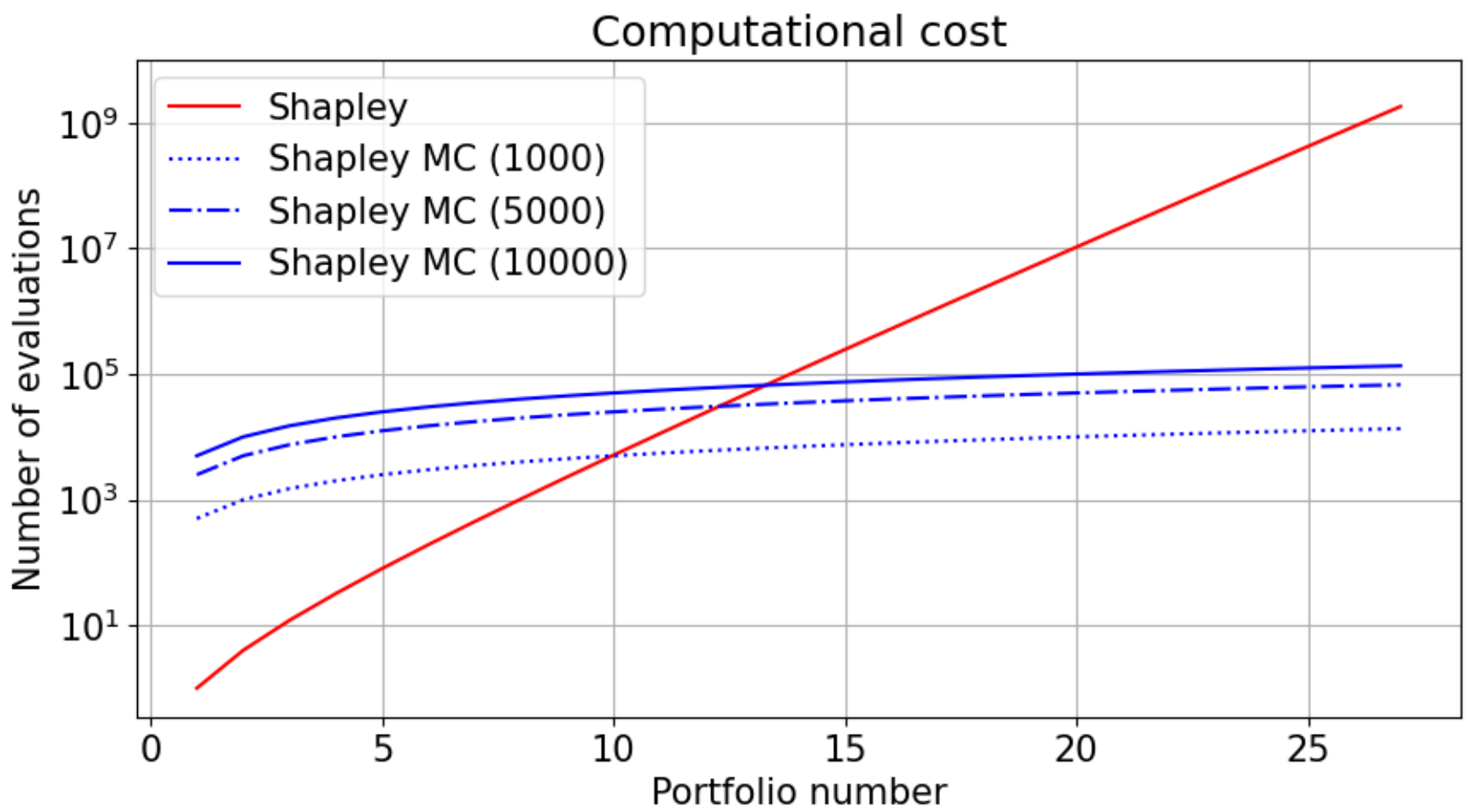

***Figure 2**: scaling of computational costs for Shapley and Shapley Monte Carlo allocation strategies with the number $n$ of risk units.*

## 3.3. Computing Marginals

Another relevant computational issue regards the calculation of marginal contributions. In concrete applications the risk measure $\rho$ is computed using historical or Monte Carlo simulations based on a finite set of $m$ empirical risk scenarios,

$$\begin{bmatrix} X_1 \longrightarrow \hat{X}_1 = \{x_{1,1}, \dots, x_{1,k}, \dots x_{1,m}\} \\ \vdots \\ X_i \longrightarrow \hat{X}_i = \{x_{i,1}, \dots, x_{i,k}, \dots x_{i,m}\} \\ \vdots \\ X_n \longrightarrow \hat{X}_n = \{x_{n,1}, \dots, x_{n,k}, \dots x_{n,m}\} \end{bmatrix}, \qquad \rho(X_\mathcal{S}) \longrightarrow \rho(\hat{X}_\mathcal{S}) = \rho\left(\sum_{j\in\mathcal{S}} \hat{X}_j\right) = \rho \begin{bmatrix} \sum_{j\in\mathcal{S}} x_{j,1} \\ \vdots \\ \sum_{j\in\mathcal{S}} x_{j,k} \\ \vdots \\ \sum_{j\in\mathcal{S}} x_{j,m} \end{bmatrix}. \tag{64}$$

This setting allows to exploit the additivity property by scenario to efficiently compute marginals in Eq. (19) as

$$\rho(X_\mathcal{S} + X_i) \longrightarrow \rho(\hat{X}_\mathcal{S} + \hat{X}_i) = \rho\left(\sum_{j\in\mathcal{S}} \hat{X}_j + \hat{X}_i\right) = \rho \begin{bmatrix} \sum_{j\in\mathcal{S}} x_{j,1} + x_{i,1} \\ \vdots \\ \sum_{j\in\mathcal{S}} x_{j,k} + x_{i,k} \\ \vdots \\ \sum_{j\in\mathcal{S}} x_{j,m} + x_{i,m} \end{bmatrix}. \tag{65}$$

This approach avoids expensive recalculations of the risk measure $\rho$ provided one has access to risk units' values by scenario.

# 4. Numerical Results: Toy Cases

In this section we provide different numerical tests on the allocation strategies introduced in Sec. 2.3 with the aim to explore their characteristics and behavior. The tests are based on a simplified framework in which the risk units follow a normal distribution. In this context the most common risk measures and allocation strategies may be computed analytically, therefore they can be considered as benchmarks to test the consistency of the corresponding numerical calculations.

## 4.1. Monte Carlo Convergence Analysis

In this section we aim to test the convergence of the numerical simulation of the allocation strategies to their expected values. To this scope we assume 10 risk units following 10 independent standard normal distributions,

$$n = 10, \qquad X_i \sim \mathcal{N}(0,1)\ \forall i, \qquad X = \sum_{i=1}^{n} X_i \sim \mathcal{N}(0, \sqrt{n}), \tag{66}$$

where $\mathcal{N}(\mu, \sigma)$ represents a normal distribution with mean $\mu$ and standard deviation $\sigma$. We select 10 risk units to allow both sufficiently rich statistics and analytical computation of Shapley allocation at the same time. Since the gaussian distributions are independent, the benchmark risk allocations are expected to yield an equal partition of the total risk, i.e.

$$K_i^B = \frac{\rho(X)}{n} \ \forall i. \tag{67}$$

Furthermore, since risk measures for gaussian distributions are known analytically for Standard Deviation, Value at Risk and Expected Shortfall, we get the exact values below,

- **Standard Deviation**: $\rho(X) = \mathrm{Std(X)}$, $K_i^B = \frac{1}{10}\mathrm{Std(X)} = \frac{\sqrt{10}}{10} \approx 0.316 \ \forall i$,
- **Value at Risk**: $\rho(X) = -\mathrm{VaR(X, 1\%)}$, $K_i^B = -\frac{1}{10}\mathrm{VaR(X, 1\%)} = -\frac{\sqrt{10}}{10}\Phi_{\mathcal{N}}^{-1}(1\%) \approx 0.736 \ \forall i$,
- **Expected Shortfall**: $\rho(X) = \mathrm{ES(X, 1\%)}$, $K_i^B = \frac{1}{10}\mathrm{ES(X, 1\%)} = \frac{1}{10}\frac{\sqrt{10}}{1\%}\phi_{\mathcal{N}}\left[\Phi_{\mathcal{N}}^{-1}(1\%)\right] \approx 0.843 \ \forall i$,

where $\Phi_{\mathcal{N}}$ and $\phi_{\mathcal{N}}$ are the standard normal cumulative and probability distribution functions, respectively. The challenger risk allocations $K_i^{\Pi}$ are therefore computed according to the risk allocation strategy $\Pi$, and compared to $K_i^B$. We stress that the match is asymptotic, depending on the number of $m$ gaussian samples of Eq. (66) employed to compute the risk measure $\rho_m(X)$. Consequently, we aim to verify that

$$K_{i,m}^{\Pi} \xrightarrow[m\to\infty]{} K_i^{\Pi} = K_i^B \tag{68}$$

We show in the following Figure 3, Figure 4 and Figure 5 the results for Standard Deviation, Value at Risk and Expected Shortfall, respectively.

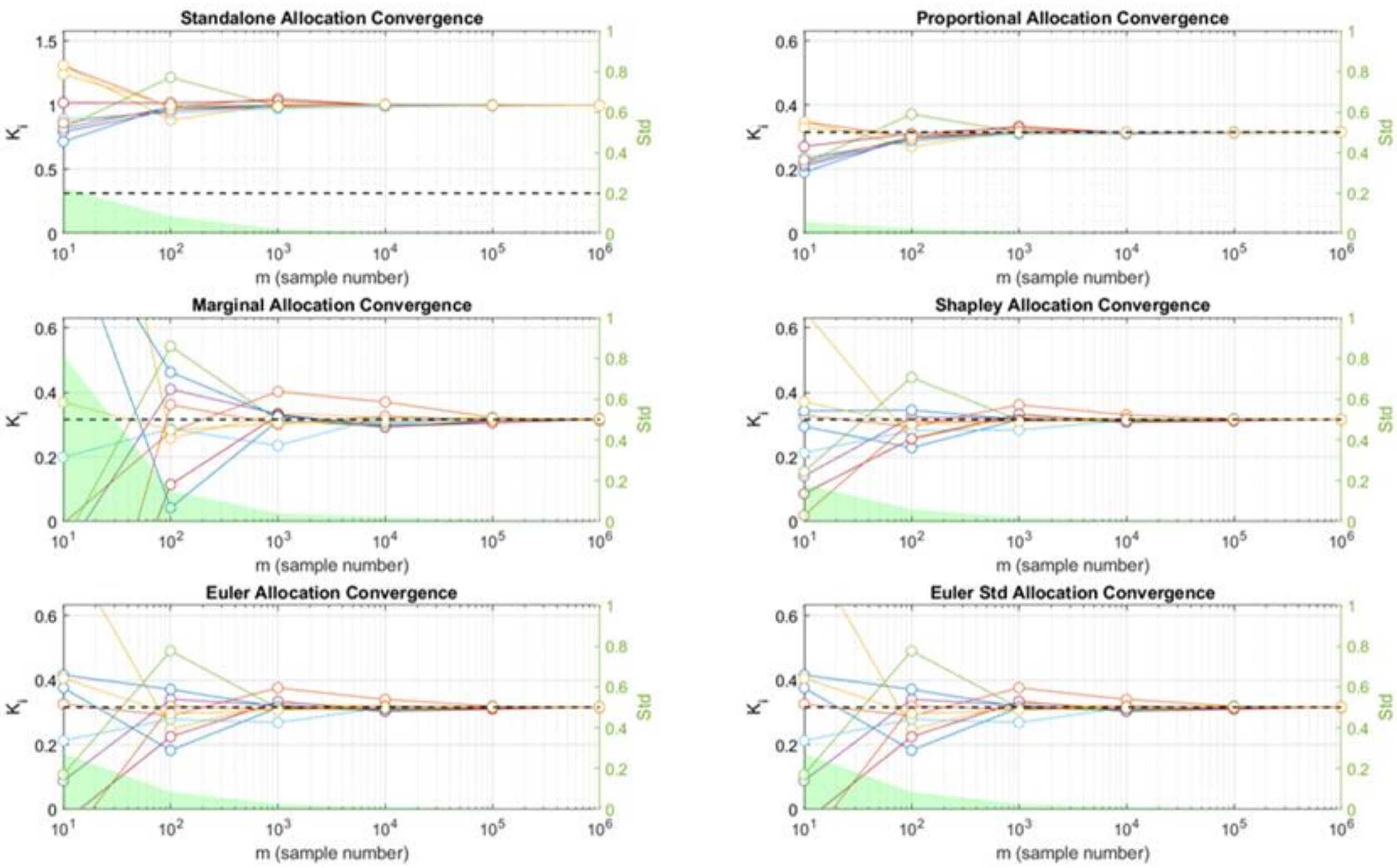

***Figure 3**: **Standard Deviation**. Convergence test as in Eq. (68) for the allocation strategies of Sec. 2.3. Horizontal axes show the log-number of samples, up to 1 million; horizontal dashed black lines represent the exact allocation; the vertical left-hand axes show the estimated allocation values. Each line represents a portfolio (10 lines are plotted); the vertical right-hand axes show the standard deviation of the estimate (green areas). The bottom right chart has been simulated using eq. (28).*

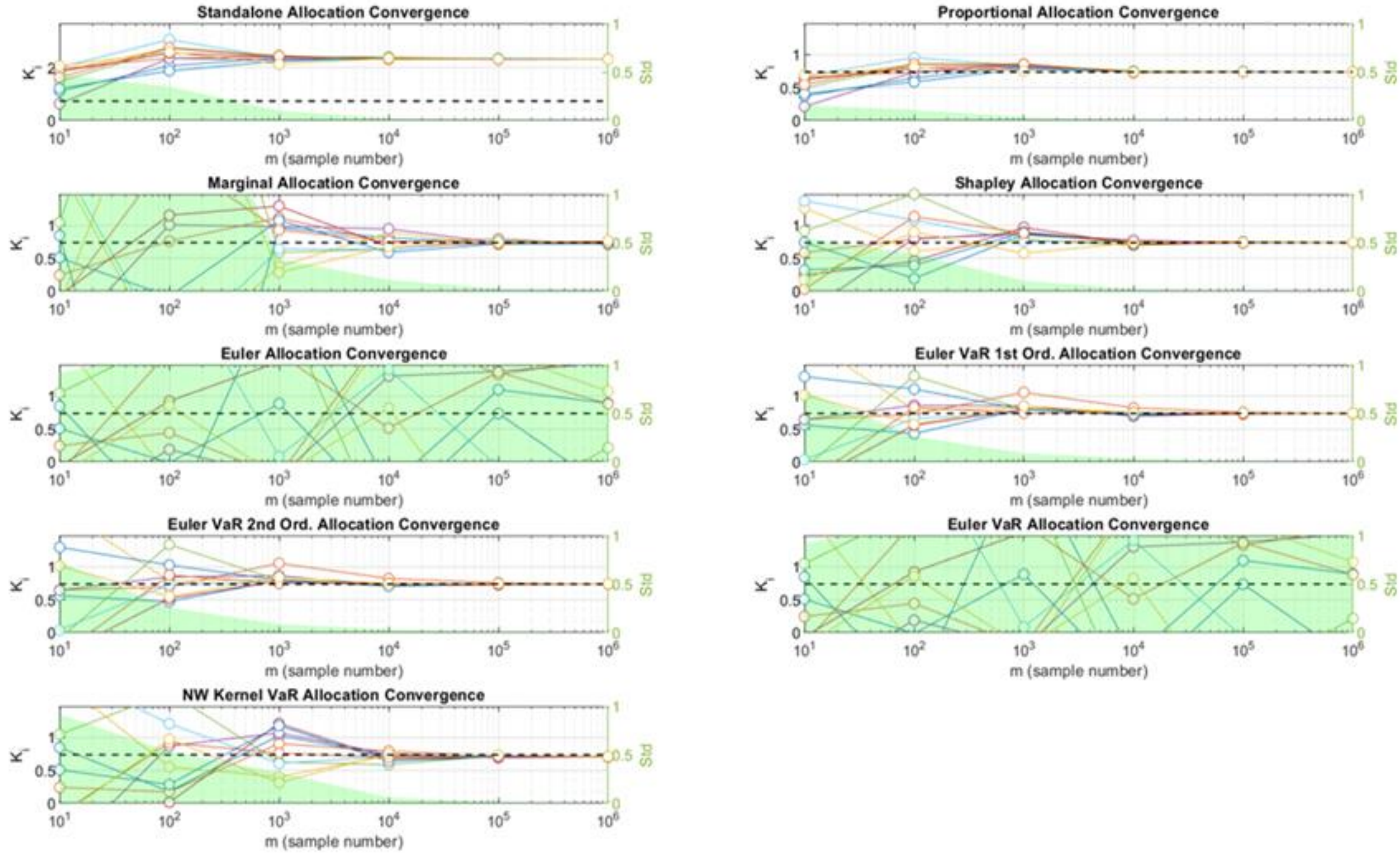

***Figure 4**: **Value at Risk**. Convergence test as in Eq. (68). Charts explanation as in Figure 3. The bottom right chart has been simulated using eq. (30).*

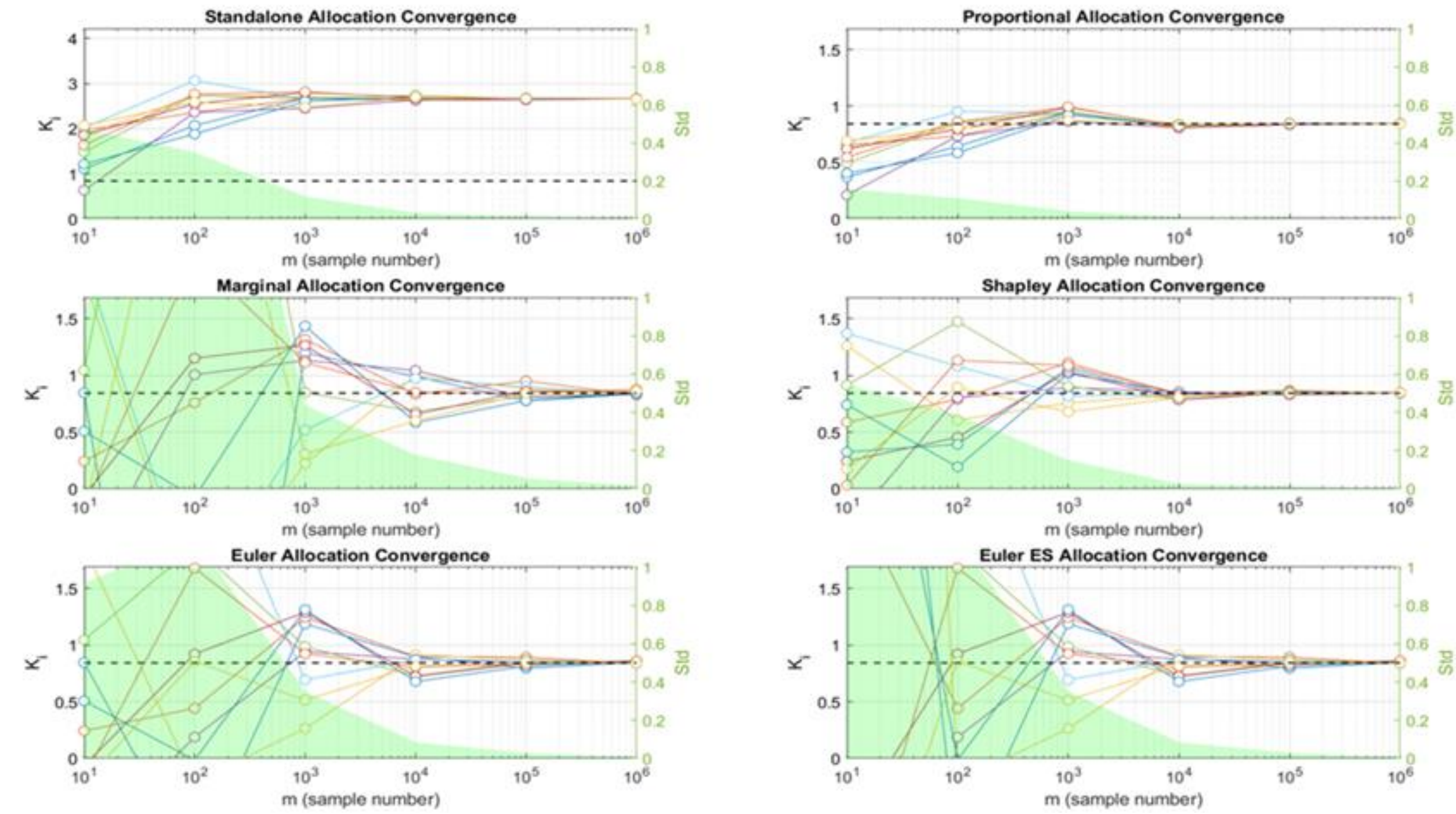

***Figure 5**: **Expected Shortfall**. Convergence test as in Eq. (68). Charts explanation as in Figure 3. The bottom right chart has been simulated using eq. (29).*

We observe that all the risk measures $\rho$ considered above the Standalone allocations converge to the wrong value, since, as discussed in Sec. 2.3.1, this strategy does not guarantee the full allocation property. In fact, since all the risk measures $\rho$ are non-linear functions of $X$, we have

$$\rho(X) = \rho(\sum_{i=1}^{n} X_i) \neq \sum_{i=1}^{n} \rho(X_i) = \sum_{i=1}^{n} K_i^{\mathrm{Sta}} = nK^{\mathrm{Sta}} \implies K^{\mathrm{Sta}} \neq \frac{\rho(X)}{n}, \tag{69}$$

in contrast with Eq. (68).

Regarding the Standard Deviation risk measure in Figure 3, we observe a fast convergence to the benchmark values for every allocation strategy. The estimation error is greatly reduced already with $10^4$ samples. The convergence is faster with respect to Value at Risk and Expected Shortfall, since they are both tail measures that typically require more samples for accurate estimation. We also observe that both general and specific Euler allocation strategies converge to the correct results, with comparable convergence patterns, consistently with the discussion of Sec. 2.3.5.

Regarding the Value at Risk measure in Figure 4, we firstly observe that both general and specific Euler strategies do not converge, consistently with the discussion of Sec. 2.3.5. In contrast, both first and second order approximated Euler strategies converge to the correct value, with negligible improvements of the second order approximation. The approximated Kernel Euler strategy converges to the expected value with a small discrepancy due to the inexact fulfil of the full allocation property, as discussed in Sec. 2.3.5, but with a slower pace with respect to the other two approximated Euler strategies. Similar convergence patterns are displayed by Shapley allocation, while proportional allocation shows the fastest convergence rate. The lowest convergence rate is displayed by the marginal allocation strategy, because of the potential numerical instabilities discussed in Sec. 2.3.3.

The results for the Expected Shortfall risk measure in Figure 5 are aligned to the ones for Value at Risk just discussed, but all the cases converge to their respective benchmark values.

Overall, we observe that Shapley allocations always converge with a good speed, inferior only to Proportional allocations. Euler does not converge in the case of VaR. We stress that Shapley allocations were computed analytically, not using Monte Carlo, since in this case the number of risk units is small enough.

## 4.2. Allocation strategies comparison

In this section we focus on the comparison between Proportional, Shapley and Euler allocation strategies discussed in Sec. 2.3.2, 2.3.4, and 2.3.5, respectively. The reason for this test is to highlight some peculiar features of the Shapley and Euler allocation strategies to be compared with Proportional allocation, that is typically considered as a benchmark because of its simplicity and diffusion. To this scope we assume 5 risk units following a 5-dimensional multivariate normal distribution $\boldsymbol{X}$,

$$n = 5, \quad \boldsymbol{X} \sim \mathcal{N}(0, C), \quad X = \sum_{i=1}^{n} X_i \sim \mathcal{N}(0, C), \tag{70}$$

where $C \in \mathbb{R}^{5\times 5}$ is the covariance matrix[6], and $\rho(X) = \mathrm{VaR}(X, 1\%)$. We select only 5 risk units since they are enough to highlight the distinct properties of Shapley and Euler allocation, for which we need multivariate (correlated) distributions. Gaussian distributions and Value at Risk measure are selected for simplicity but are not essential to our scope. Since the Proportional allocation strategy, as discussed in Sec. 2.3.2, does not consider correlations between risk units, the Proportional allocations depend only on variances. Since all the variances are equal to 1, we conclude that, in this case, $K_i^{\mathrm{Pro}} = \rho(X)/n \ \forall\, i$, and any difference between Proportional, Shapley and Euler allocations is due only to correlations. The VaR figures are computed numerically using $m = 10^6$ samples from the multivariate gaussian distribution, then Proportional, Euler and Shapley allocations are computed.

[6] Being all the variances equal to one, the covariance matrix is equal to the correlation matrix.

We provide in Figure 6 the results of this test for different covariance matrices, which represents different relationships between risk units.

| | $X_1$ | $X_2$ | $X_3$ | $X_4$ | $X_5$ |
|---|---|---|---|---|---|
| $X_1$ | 1 | -0.3 | 0.3 | 0.3 | 0.3 |
| $X_2$ | -0.3 | 1 | -0.3 | -0.3 | -0.3 |
| $X_3$ | 0.3 | -0.3 | 1 | 0.3 | 0.3 |
| $X_4$ | 0.3 | -0.3 | 0.3 | 1 | 0.3 |
| $X_5$ | 0.3 | -0.3 | 0.3 | 0.3 | 1 |

| | $X_1$ | $X_2$ | $X_3$ | $X_4$ | $X_5$ |
|---|---|---|---|---|---|
| $X_1$ | 1 | -0.3 | 0.3 | -0.3 | 0.3 |
| $X_2$ | -0.3 | 1 | -0.3 | -0.3 | -0.3 |
| $X_3$ | 0.3 | -0.3 | 1 | -0.3 | 0.3 |
| $X_4$ | -0.3 | -0.3 | -0.3 | 1 | -0.3 |
| $X_5$ | 0.3 | -0.3 | 0.3 | -0.3 | 1 |

| | $X_1$ | $X_2$ | $X_3$ | $X_4$ | $X_5$ |
|---|---|---|---|---|---|
| $X_1$ | 1 | -0.1 | 0.2 | 0.3 | 0.3 |
| $X_2$ | -0.1 | 1 | -0.1 | -0.5 | -0.1 |
| $X_3$ | 0.2 | -0.1 | 1 | 0.5 | -0.3 |
| $X_4$ | 0.3 | -0.5 | 0.5 | 1 | 0.1 |
| $X_5$ | 0.3 | -0.1 | -0.3 | 0.1 | 1 |

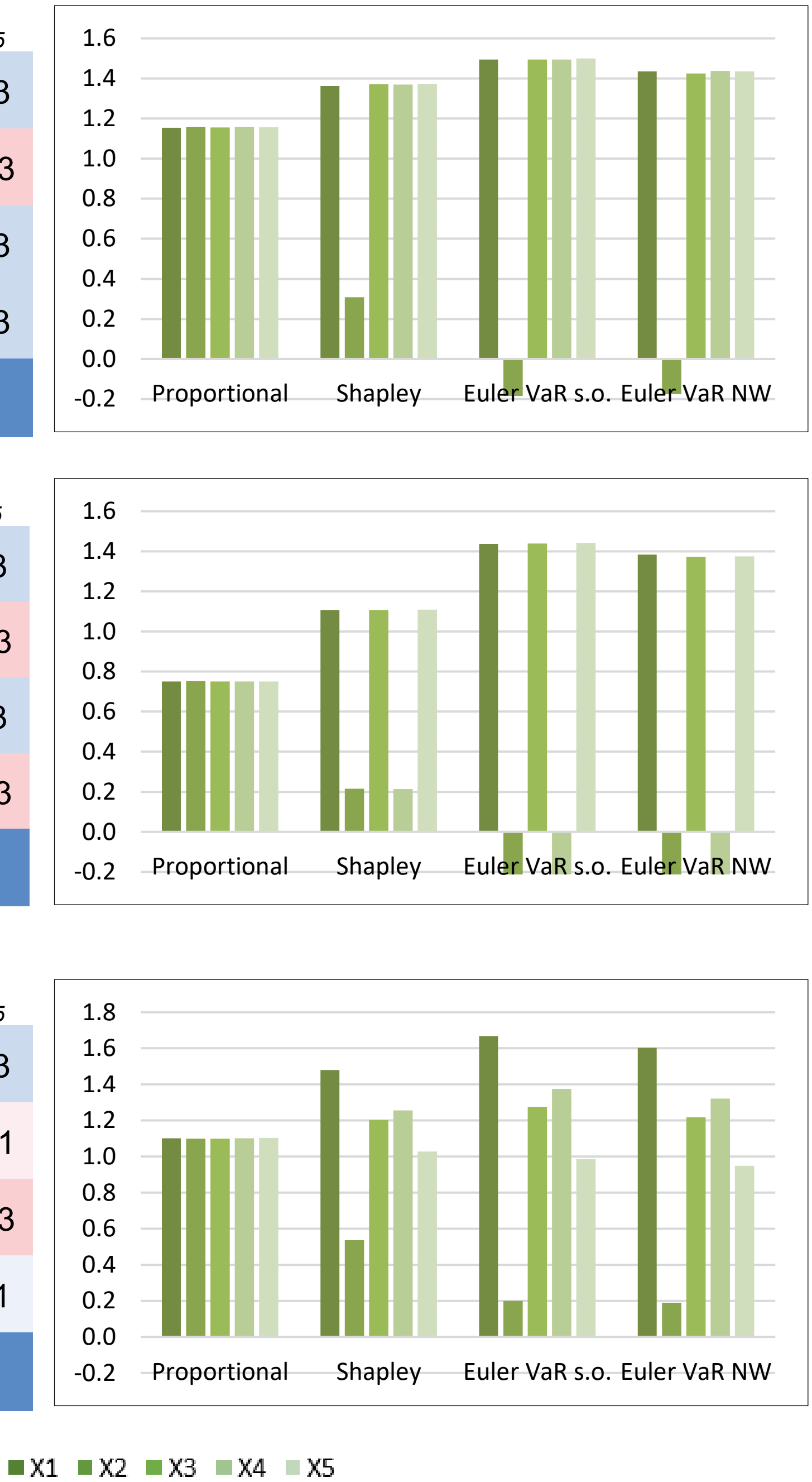

***Figure 6**: Comparison between Proportional, Shapley and Euler allocation strategies for five risk units with multivariate Gaussian distribution with different covariance matrices (reported on the left-hand side) and* $\rho(X) = VaR(X, 1\%) = 5.78, 3{,}75{,}5.51$ *for top, middle and bottom charts, respectively.*

In the first case (Figure 6, top), all the correlations have the same value ($+30\%$) but risk unit #2, which is negatively correlated with all the others ($-30\%$). Hence, risk unit #2 contributes to the overall reduction of the Value at Risk, since it partially compensates the dynamics of the other risk units (i.e.

#2 hedges the risk). This behaviour is correctly captured by the Shapley strategy, which strongly reduces the risk allocated to risk unit #2 ($-73.4\%$) and equally increases all the others ($\approx +18.4\%$) since they have equal correlations with respect to Proportional allocations.
In the second case (Figure 6, middle) we add a second risk unit #4 with the same negative correlations ($-30\%$), and leave everything else unchanged, thus increasing the hedging effect. Consistently, the Shapley strategy reduces the risk allocated to both risk units #2 and #4 ($\approx -71.4\%$) and increases all the others ($\approx +47.7\%$) with respect to the Proportional allocations.
Finally, in the third case (Figure 6, bottom) we assign a random correlation structure to each risk unit in the range $[-30\%; 30\%]$. Now Shapley allocations assume distinct values, however risk unit #2 still receives the lowest allocation, being the only one with negative correlations with respect to all the others.

For what concerns the Euler strategy, we apply two versions, namely the second order approximation for Value at Risk and the Nadaraya-Watson Kernel method (see Sec. 2.3.5). We observe that the two Euler approaches display a perfect match for all the cases taken into consideration in Figure 6, with a strong numerical consistency. Furthermore, the overall pattern of portfolio allocations aligns closely with that observed under the Shapley-based strategy, particularly in the third scenario (Figure 6, bottom). However, in the first two scenarios, both Euler approaches attribute small negative allocations to portfolios characterized by negative correlations. This contrasts with the Shapley strategy, which assigns small but strictly positive contributions to the same portfolios. These results suggest that the Euler method is more susceptible to generating negative allocations in the presence of diversification effects, highlighting a key methodological difference between the two approaches and a potential issue for Euler strategy, for which we refer to Sec. 2.4.

## 4.3. Shapley Monte Carlo

In this section we focus on the details of the Shapley Monte Carlo simulation, which, as discussed in Sec. 3.1, is the only viable solution even for a moderate number of risk units. To this scope we assume the same setting adopted in Sec. 4.1, Eq. (66), with $n = 20$ and $\rho(X) = \mathrm{VaR}(X, 1\%)$. In this case we have $2^{19} = 524{,}288$ coalitions in the exact Shapley formula, Eq. (20), and $19! = 1.2 \times 10^{17}$ permutations in the more symmetric form of Eq. (60). Hence, we resort to Shapley Monte Carlo allocation of Eq. (61) using Algorithm 1 described in Sec. 3.1.

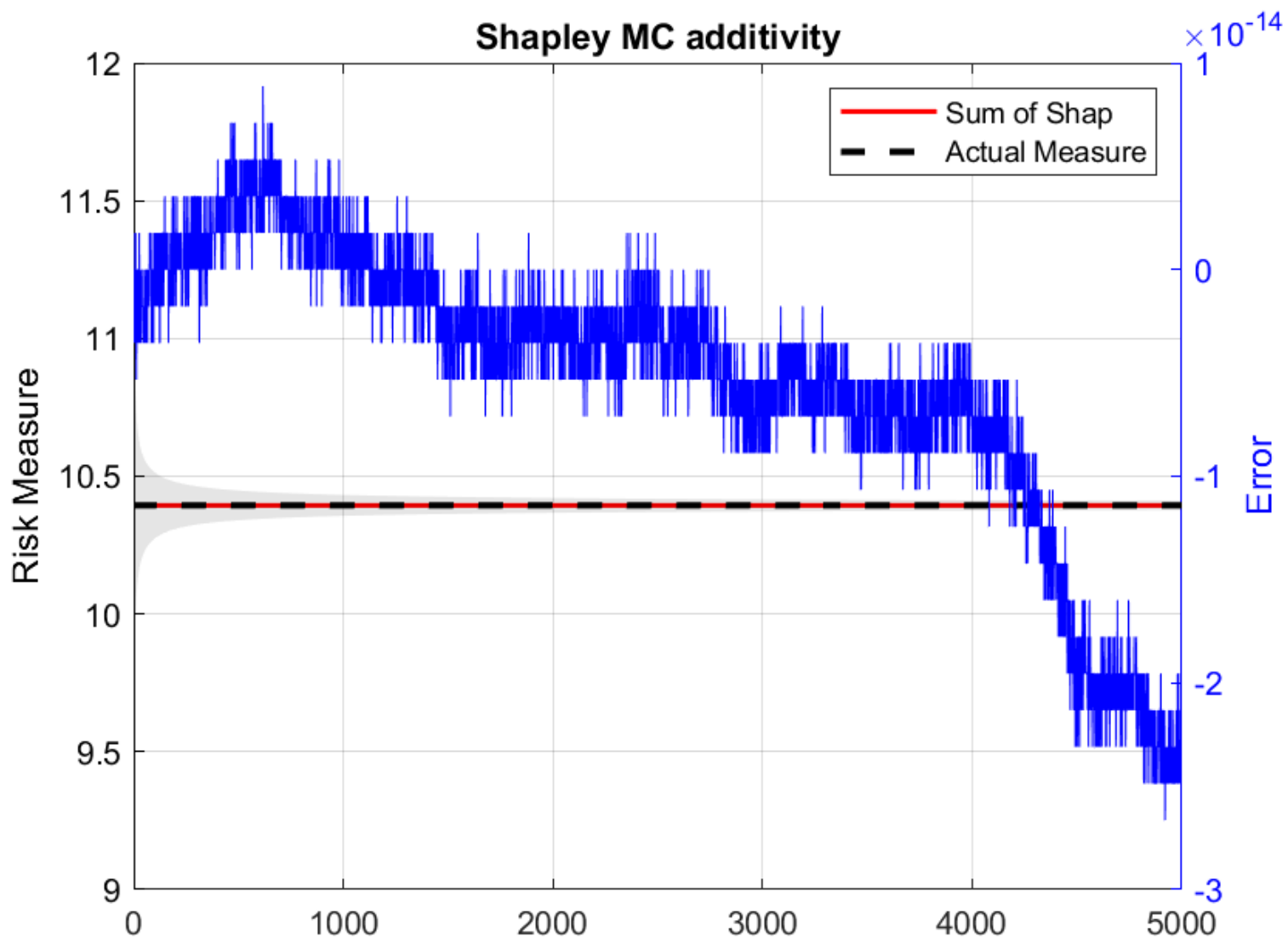

***Figure 7**: Difference (blue line, r.h.s. axis) between the sum of Shapley Monte Carlo values (dashed red line, l.h.s. axis) and the actual overall measure (black constant line, l.h.s. axis). The grey area (l.h.s. axis) is the Monte Carlo error of the sum of Shapley Monte Carlo values.*

First of all, we check empirically that Algorithm 1 respects the full allocation property of Eq. (63) for each Monte Carlo sample, as mathematically proved in App. 8.3. In fact, Figure 7 shows that, irrespective of the number of samples, the sum of the Shapley Monte Carlo allocations $K_i \approx 0.520$ is always equal to the total risk measure $\rho(X, 1\%) = 10.404$. Finally, we show in Figure 8 the Monte Carlo convergence results. Similarly to Sec. 4.1, the distribution of the portfolio being known in advance, we can compare the estimates with the true value of the allocation. Since each portfolio shares the same distribution, we expect that asymptotically the allocation of each portfolio is the same. This is what we observe, and a good convergence level is obtained already with 5,000 antithetical samples. However, we stress that the number of MC samples needed for an acceptable convergence depends on the risk measure.

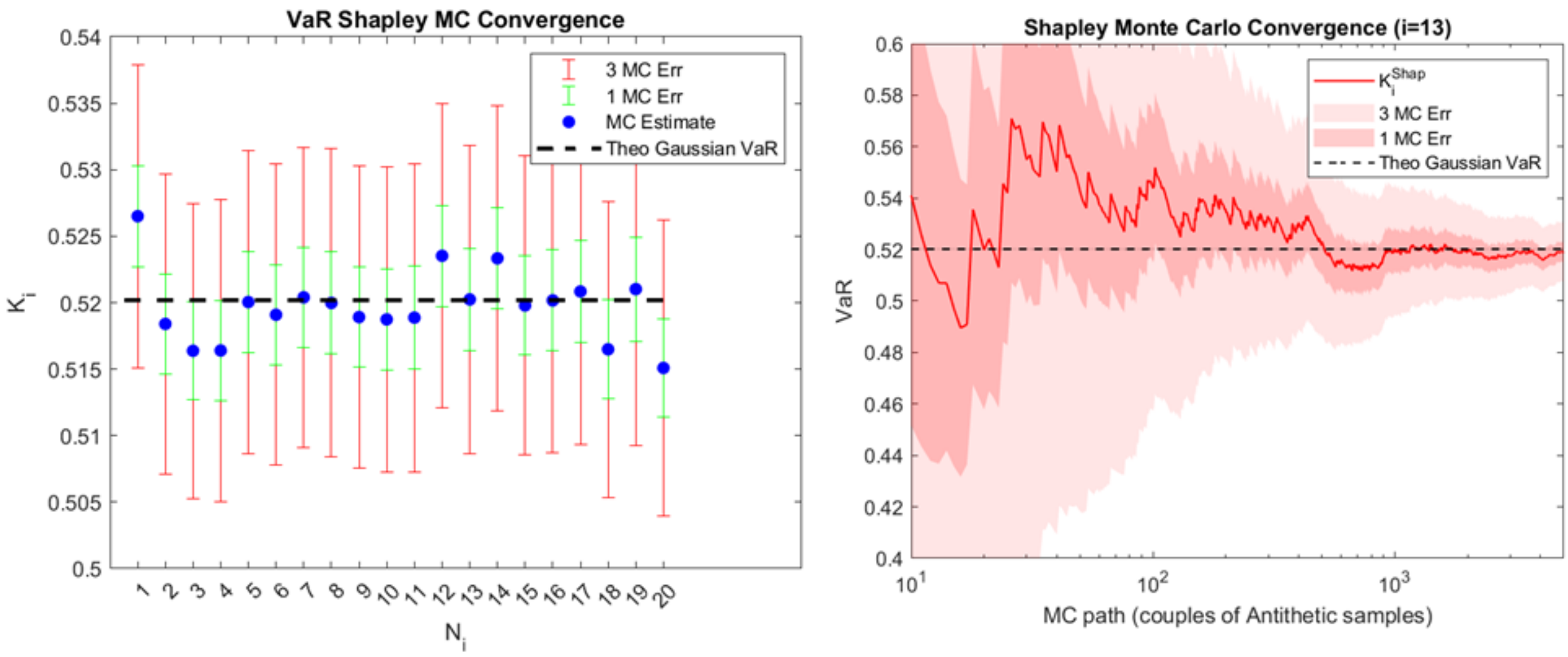

***Figure 8**: Left hand side chart: Monte Carlo Shapley value estimates for each portfolio, together with the expected theoretical value (horizontal black dashed line) and the whisker of the Monte Carlo errors, corresponding to 5,000 antithetic samples. Right hand side chart: convergence pattern of the Monte Carlo estimates for portfolio #13 together with Monte Carlo errors (1 and 3 stdev), with respect to the theoretical value (black dashed line).*

## 4.4. Negative Shapley Allocation

We discussed in Sec. 2.4 that negative Shapley values are not excluded a priori. We show in Figure 9 two typical simplified situations in which negative values may occur, assuming 5 risk units following a 5-dimensional multivariate normal distribution and $\rho = \mathrm{VaR}(1\%)$. The framework is the same adopted in Sec. 4.2, Eq. (70).

In the top row of Figure 9 we present a situation in which one portfolio (portfolio 1) displays high negative correlation with the other ones, overall contributing to the reduction of the risk measure in the coalitions. This results in a negative Shapley value for portfolio 1. The magnitude of such Shapley value is the lowest one, consistently with the observations of Sec. Sec. 2.4.

The second case depicted in the second row of Figure 9 outlines the case in which the correlation structure is almost homogeneous through portfolios, but the samples of portfolio 1 are multiplied by a scaling factor equal to 0.65. In this way we lower the relevance of portfolio 1 then computing the risk measure, simulating the common situation in which one or more portfolios of the firm are negligible in terms of overall fair value and hence in terms of contribution to the Value at Risk. The resulting Shapley value is negligible and negative. The negative sign is due to the fact that, because of the minor relevance of the portfolio, each component of the Shapley value sum of Eq. (17) has the form $\rho(X_{\mathcal{S}}) - \rho\left(X_{\mathcal{S}\setminus\{1\}}\right) \approx 0$ for almost every $\mathcal{S}$, leading to $K_i^{\mathrm{Sha}} \gtrsim 0$ or (as in the presented example) $K_i^{\mathrm{Sha}} \lesssim 0$. The proportional allocation strategy, conversely, does not exhibit any negative allocation, as expected.

| | $X_1$ | $X_2$ | $X_3$ | $X_4$ | $X_5$ |
|---|---|---|---|---|---|
| $X_1$ | 1 | -0.5 | -0.1 | -0.5 | -0.8 |
| $X_2$ | -0.5 | 1 | 0.2 | 0.3 | 0.3 |
| $X_3$ | -0.1 | 0.2 | 1 | 0.5 | -0.3 |
| $X_4$ | -0.5 | 0.3 | 0.5 | 1 | 0.1 |
| $X_5$ | -0.8 | 0.3 | -0.3 | 0.1 | 1 |

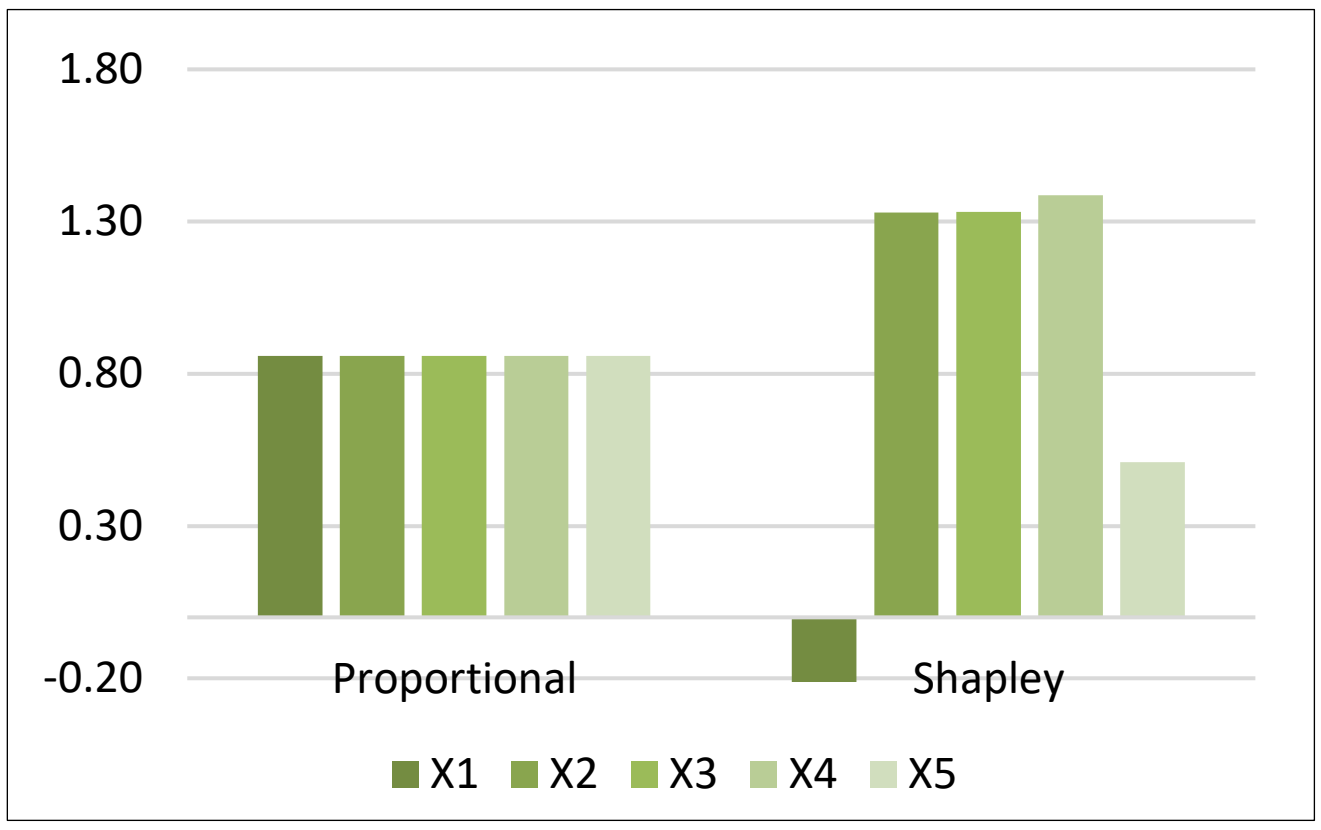

| | $X_1$ | $X_2$ | $X_3$ | $X_4$ | $X_5$ |
|---|---|---|---|---|---|
| $X_1$ | 1 | -0.3 | 0.3 | -0.3 | 0.3 |
| $X_2$ | -0.3 | 1 | -0.3 | -0.3 | -0.3 |
| $X_3$ | 0.3 | -0.3 | 1 | -0.3 | 0.3 |
| $X_4$ | -0.3 | -0.3 | -0.3 | 1 | -0.3 |
| $X_5$ | 0.3 | -0.3 | 0.3 | -0.3 | 1 |

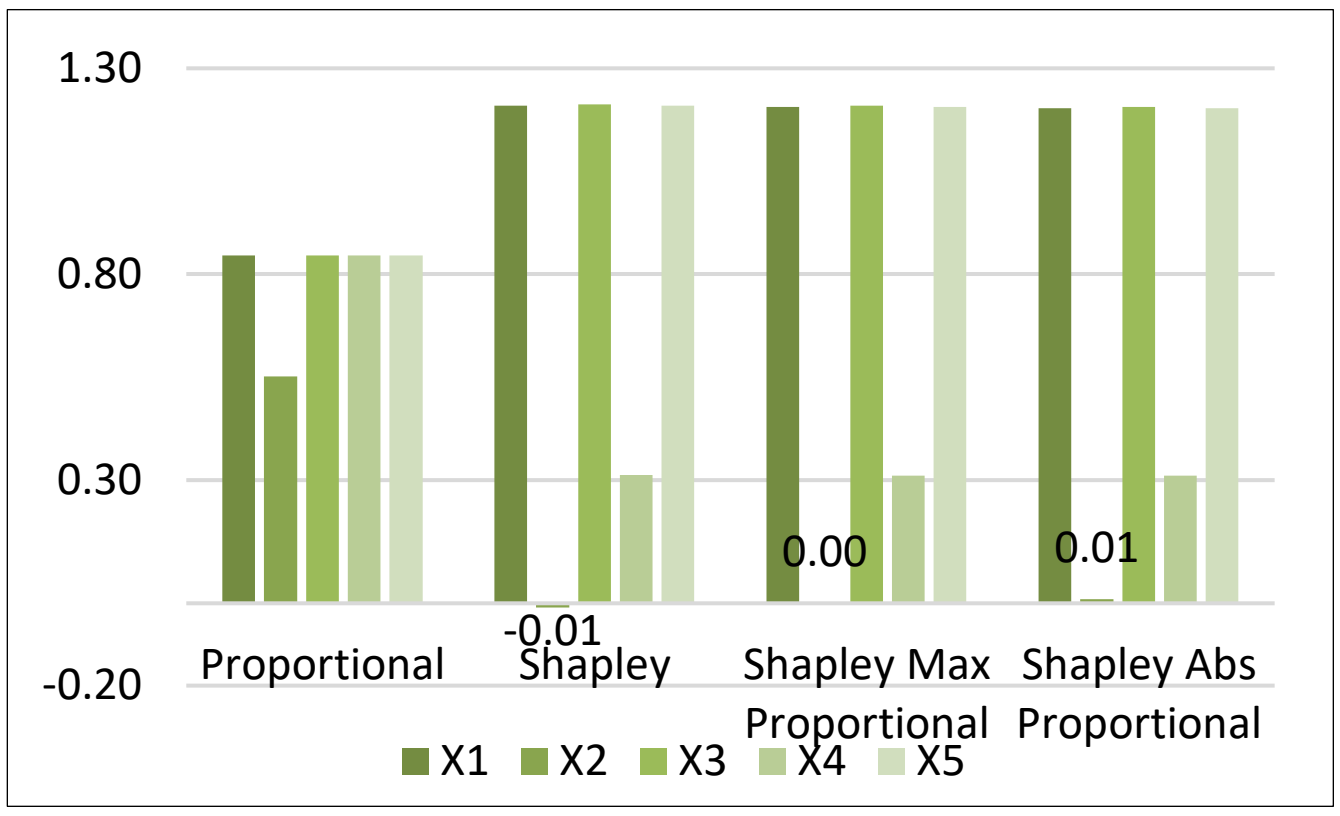

***Figure 9:*** *Comparison between Proportional and Shapley allocation strategies for five risk units with multivariate Gaussian distribution with different covariance matrices (reported on the left-hand side) leading to a negative Shapley value for portfolio 1. For the second case, the samples of portfolio 1 are multiplied by a scaling factor equal to 0.65.* $\rho(X) = VaR(X, 1\%) = 4.3, 3.9$ *for top and bottom charts, respectively.*

Finally, we show in Figure 10 the results of the application of the two alternative workarounds to remove the negative Shapley values from the affected portfolios, namely the Shapley Maximum Proportional (top right chart) and Shapley Absolute Proportional (bottom right chart) described in Sec. 2.4. We applied both to the same portfolio shown in the top left chart of Figure 9, whose covariance matrix is also reported in Figure 10. For both the approaches, the reproportioning leads to a reduction of the Shapley values of all the portfolios. In Shapley Maximum Proportional the former negative Shapley value (corresponding to portfolio 1) is set to zero, while in Shapley Absolute Proportional it becomes positive, with a magnitude which is comparable with the former magnitude. The choice between the two approaches should be driven by the specific nature of the problem, in particular by the risk measure involved.

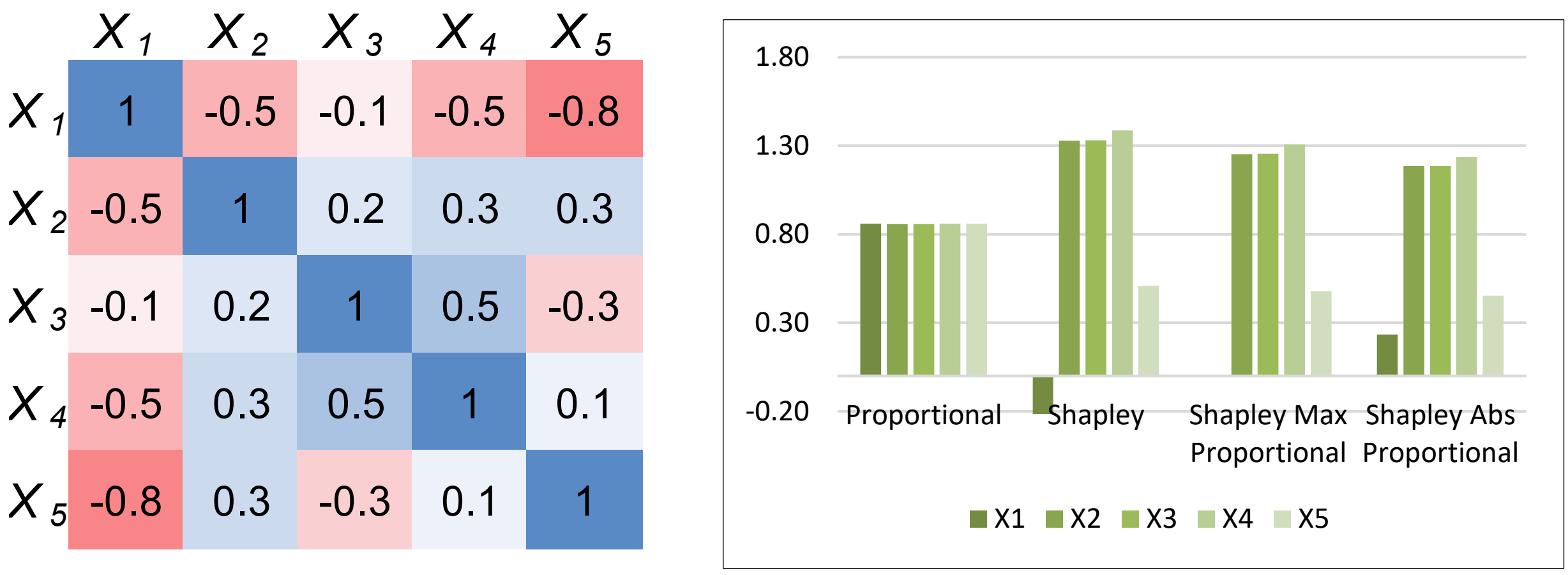

***Figure 10:*** *Comparison between Proportional and Shapley allocation strategies for five risk units with multivariate Gaussian distribution with the covariance matrix represented in the first column. Second column shows the results of the application of Shapley Max Reproportioned and Shapley Abs Reproportioned, respectively.* $\rho(X) = VaR(X, 1\%) = 4.3$.

# 5. Numerical Results: Realistic Cases

In this section we provide the application of Shapley and Proportional allocation strategies to realistic trading portfolios. The first application, in Sec. 5.1, considers the value at risk measure, adopted under the Basel 2.5 market risk capital framework (CRR1, 2013, and CRR2, 2019). The second application, in Sec. 5.2, considers the equity component of the sensitivity-based (SBA) risk measure adopted under the Simplified Approach allowed by the Fundamental Review of Trading Book (FRTB) market risk capital framework (CRR3, 2024).

## 5.1. Market Risk Capital Allocation Under Basel 2.5

Value at Risk (VaR) is a widely adopted metric for assessing the potential loss of a portfolio. It serves various purposes, including managerial decision-making, regulatory compliance, and financial reporting. In particular, the regulatory market risk capital charge under the Basel 2.5 market risk framework (CRR1, 2013, and CRR2, 2019) is based on the calculation of Value at Risk (VaR) under the current market scenario (last 250 historical risk factors' observations), of Stressed Value at Risk (sVaR) under a stressed market scenario (the worst period including 250 historical risk factors' observations for the current portfolio), and of the Incremental Risk Charge (IRC), capturing losses arising from defaults and credit migrations of entities within the portfolio. Although the Basel 2.5 framework is replaced by the FRTB framework since January 2027, the VaR measure remains however broadly adopted in the financial institutions for a wide range of applications and managerial measurement of the portfolio riskiness. The full Basel 2.5 capital charge formula is reported in Sec. 8.4. In this section we consider the following risk measure to be allocated in each of the underlying portfolios is

$$\rho(X, X') = -[\mathrm{VaR}(X, 1\%) + \mathrm{sVaR}(X', 1\%)], \tag{71}$$

where we excluded the IRC component because of the significant computational challenges posed by the Monte Carlo simulations required for IRC allocation. A potential workaround to include the IRC is described in Sec. 2.6.

We apply the allocation strategies introduced in Sec. 2.3 to a realistic trading book encompassing hundreds of thousands of financial instruments across all asset classes and organized into three hierarchical levels of trading desks. Specifically, we compare the following allocation strategies:

- Proportional allocation strategy of Sec. 2.3.2;
- Shapley allocation strategy, in its original version of Sec. 2.3.4, handling the negative allocation values as described in Sec. 2.4 and taking into account the different organizational layers as illustrated in Sec. 2.5 (Multi-level Shapley). We point out that the lowest level comprises 24 trading desks, necessitating the use of a Monte Carlo approach to compute the Shapley value. The intermediate and highest levels comprise 12 and 4 trading desks respectively, allowing an exact computation of the Shapley value.
- Euler allocation strategy (Sec. 2.3.5) for value at risk, specifically the second order approximation and the one based on Nadaraya-Watson Kernel.

In App. 8.4 we provide the details of the trading book hierarchy, the standalone VaR and sVaR measures for each trading desk (normalized to total $\rho(X)$ in Eq. (71)), and the corresponding historical correlation matrices that underpin the results of the allocations. In the following we present the key outcomes of the allocation strategies to the three hierarchical levels reported in App. 8.4.

In the top panel of Figure 11 we provide the outcomes for the first hierarchical level (4 trading desks). The results are determined by two key factors: the standalone portfolio risk, which is captured by each allocation, and historical correlations, which are accounted exclusively by the Shapley and Euler allocations. These factors are detailed in App. 8.4, Table 6, where we observe that PTF_4 has the largest standalone risk measure driven by sVaR. As a result, the Proportional allocation assigns the largest share of the overall measure to this portfolio. However, the correlation structure reveals that PTF_4 exhibits the largest negative correlations pattern for sVaR with the most relevant portfolios (PTF_1 and PTF_3) hedging most of the 8 possible coalitions and reducing Shapley with respect to Proportional allocation. For what concerns the Absolute Shapley allocation results, they differ from plain Shapley allocation because of PTF_2, which displays negligible negative allocation figure. Hence, the adjustment introduced by the Absolute Shapley technique is very small. In the top panel of Figure 11 we also compare the Shapley allocation with the Bottom-up Shapley approach, obtained starting from the Shapley allocation of the second layer as discussed in Sec. 2.5. Finally, we include the allocation results coming from the Euler-based allocation principle, tailored for allocation problems in which the risk measure is the value at risk, namely the one based on a second-order linear approximation of the expected value, and one Nadaraya-Watson Kernel (see Sec. 2.3.5). Both Euler-based allocations differ significantly from other allocation principles and do not closely approximate the theoretical Euler method.

We provide in the middle and bottom panels of Figure 11 the results for the second and third hierarchical levels, consisting in 12 and 48 trading desks, respectively (see App. 8.4). We still adopted an exact Shapley approach for the second level, since the number of portfolios is manageable. Conversely, for the third level we had to adopt the Monte Carlo Shapley estimate[7]. We observe overall the same behavior already described for the first layer, but now reflected in the finer portfolio structure, with a few peculiarities worth pointing out. First, lower-level portfolios allow less aggregation and diversification and therefore may lead to more negative allocations, that determine more differences between Shapley and Shapley Absolute allocations. Regarding negative allocations, we observe that the signs of allocation amounts are consistent between Shapley and Euler based allocations, as expected. We finally point out that for lower organizational levels we introduced the Shapley Proportional Top Down (PTD) allocation discussed in Sec. 2.5.1.

[7] We highlight that we applied the Absolute Shapley adjustment after the Monte Carlo simulation in which the plain Shapley technique is used.

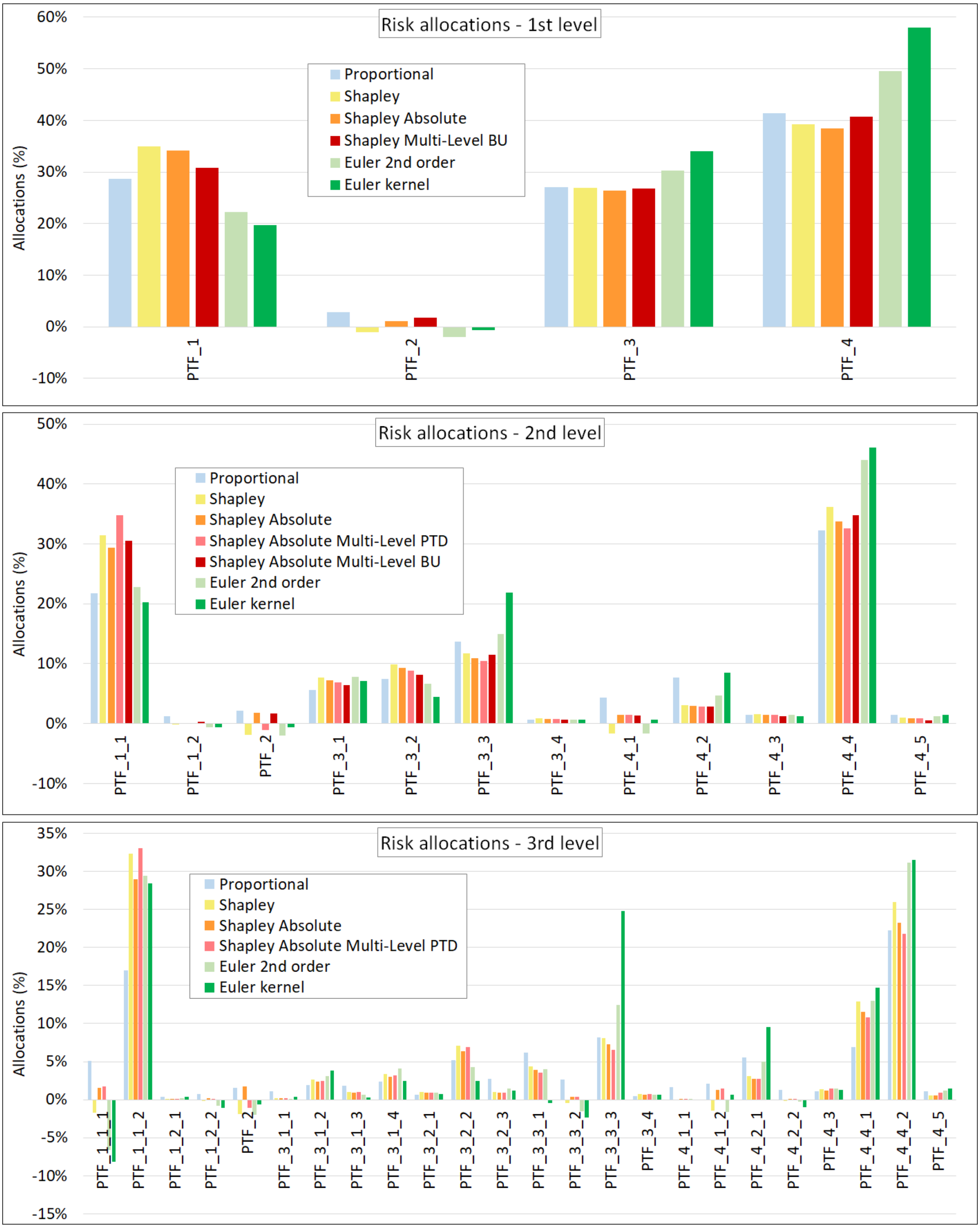

***Figure 11**: Allocations of VaR+sVaR figures, normalized to total VaR+sVaR $\rho(X)$ in Eq. (71), for the first (top panel), second (mid panel) and third (bottom panel) hierarchical levels (see App. 8.4). Monte Carlo Shapley allocations in the bottom panel are computed using 5,000 pairs of antithetical MC scenarios. The MC errors (three standard deviations) for Shapley allocations are not visible thanks to the good convergence level.*

In Figure 12 we provide a breakdown of the data presented in Figure 11 for the first hierarchical level. Specifically, we show the 8 components of the Shapley allocation formula of Eq. (17) for each portfolio, corresponding to the 8 possible coalitions, along with the relative cumulative patterns. Among the 8 coalitions, the Marginal and Standalone terms (see discussion in Sec. 2.3.4) emerge as the most significant contributors to the Shapley allocation. Although PTF_4 is dominant in terms of the standalone figures, its marginal contribution is negligible. This is due to the netting between VaR and sVaR due to the specific correlation structure of these portfolios discussed above.

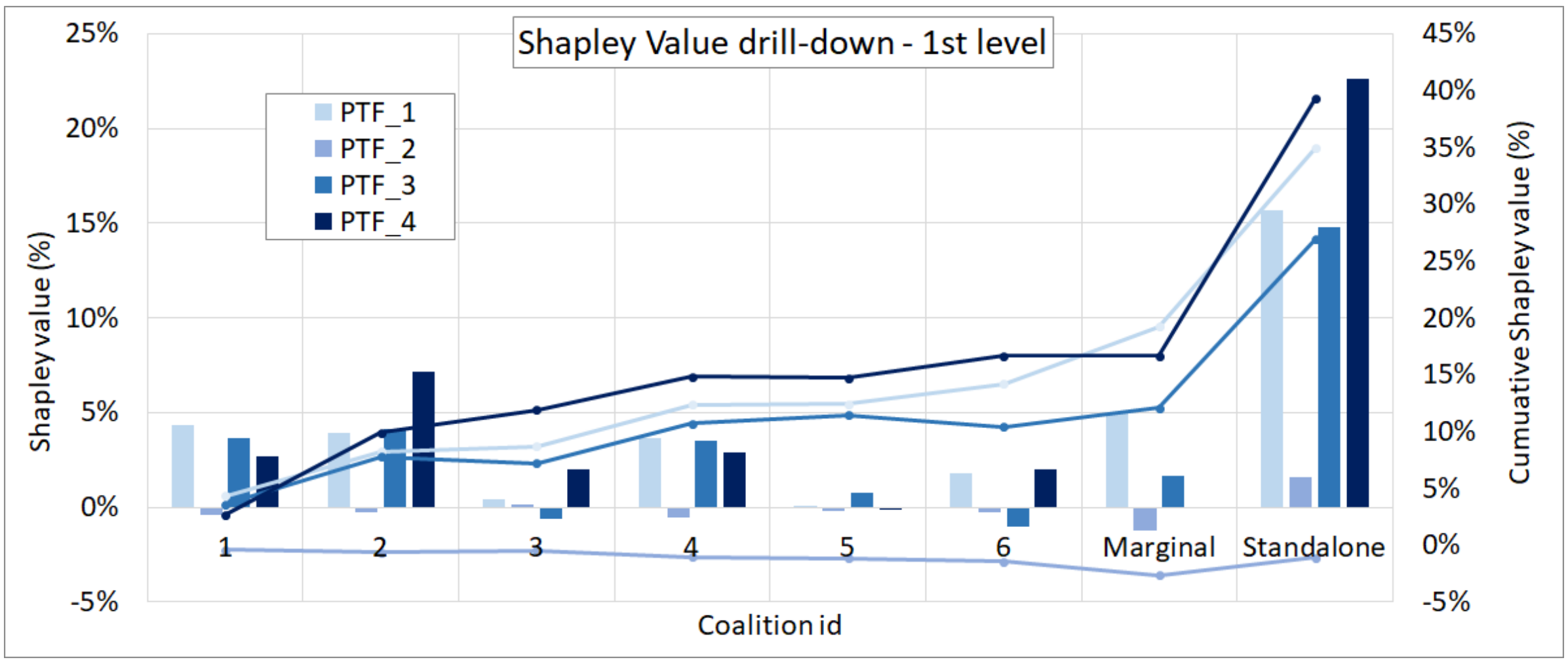

***Figure 12**: Breakdown of the Shapley allocations represented in Figure 11 (top panel). Solid line (right hand axis): cumulative sums of the components of Shapley allocations. The final points correspond to the Shapley allocations shown* Figure 11.*Vertical bars (left hand axis): Shapley allocations for each coalition (PTF_2 is not always visible since its contribution is very small). The labels of the first six coalitions is arbitrary. All the amounts are normalized to total VaR+sVaR* $\rho(X)$ *in Eq. (71).*

In Figure 13 we provide a drill-down for the Shapley allocation figures of the four portfolios leading to the largest contribution at the 2nd portfolio level according to Figure 11 (mid panel).

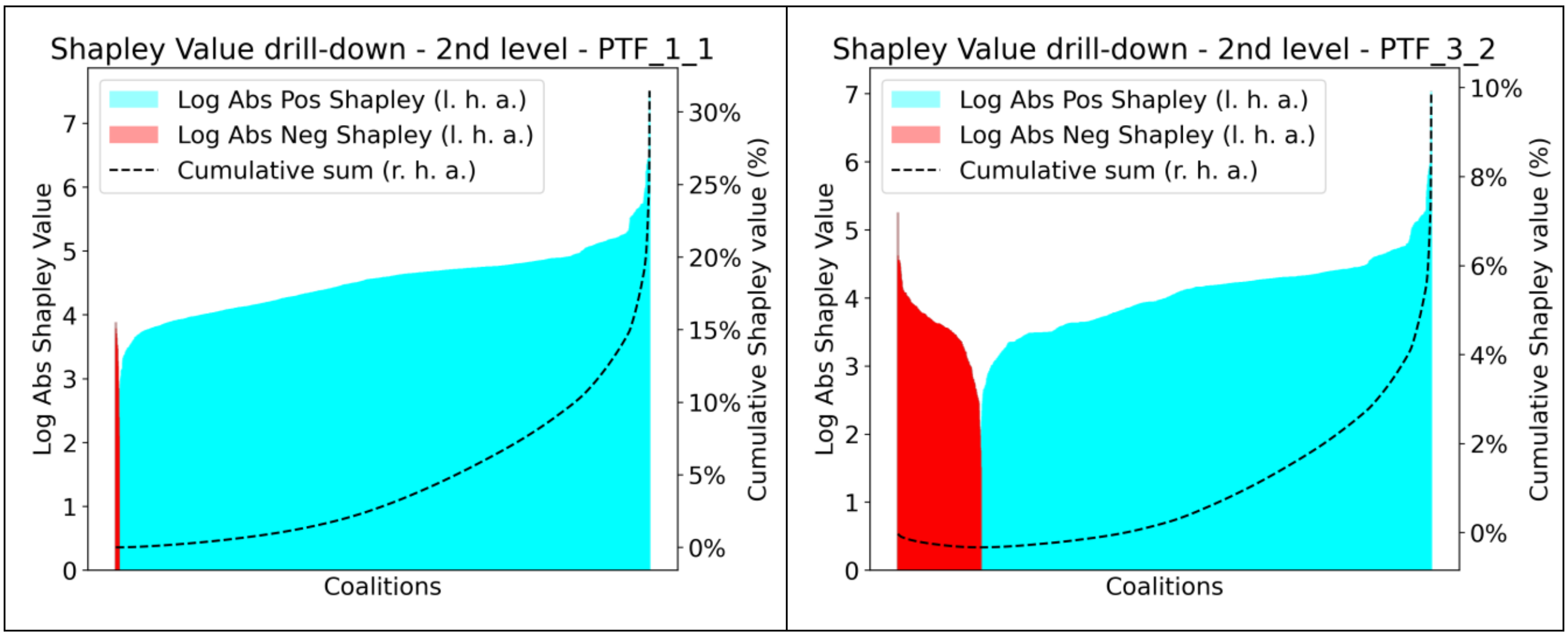

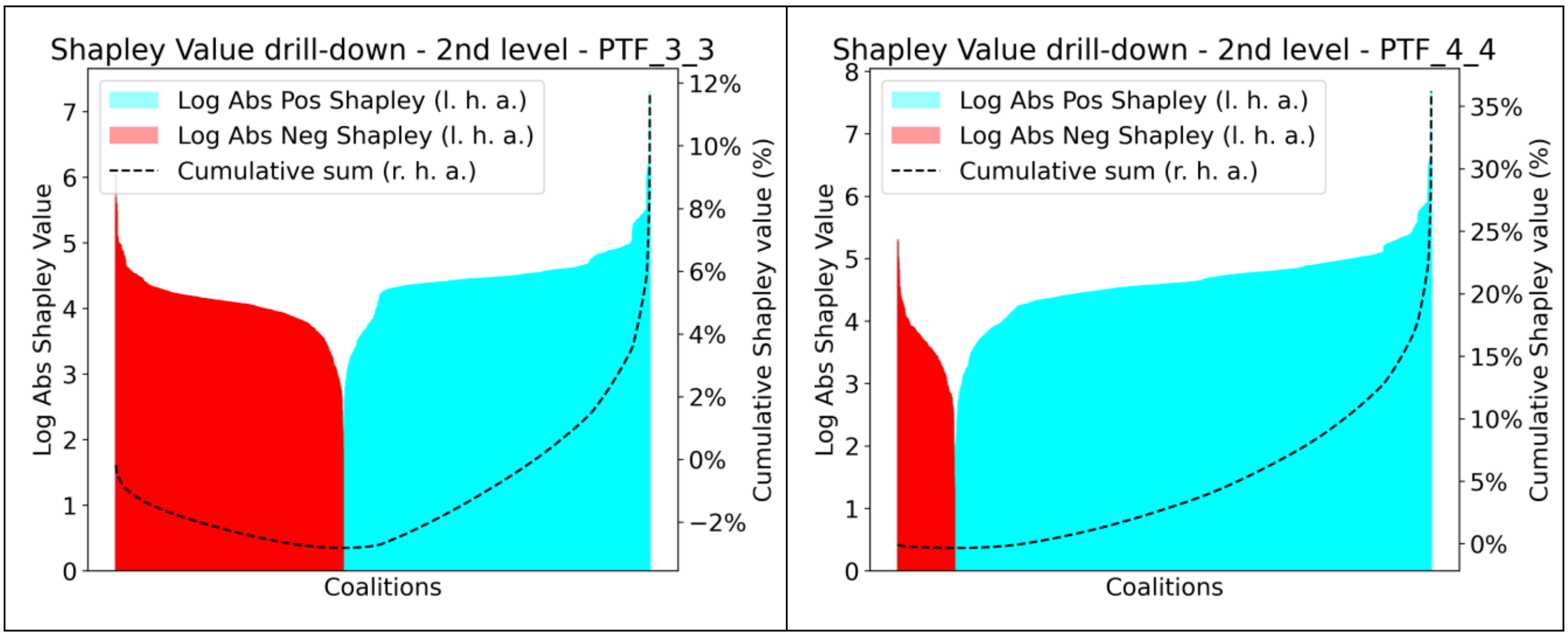

***Figure 13**: Breakdown of the Shapley values of the four portfolios leading to the largest contribution at the 2nd portfolio level according to Figure 11 (mid panel). The 2048 possible Shapley coalitions are displayed along the horizontal axis, ordered by the corresponding Shapley amounts from the largest negative to the largest positive. Red and cyan bars (left hand axis): $log_{10}$ of the absolute values of each Shapley value. Originally negative amounts are reported in red, positive ones in cyan. The logarithmic scale highlights differences across several orders of magnitude. Solid red line (right hand axis): cumulative sum of Shapley allocation components, normalized to total VaR+sVaR $\rho(X)$ in Eq. (71). The sharp increases observed in the cumulative curve are driven by a small number of dominant Shapley values, typically associated with Proportional and Marginal coalitions (see discussion in Sec. 2.3.4).*

This convergence of Shapley Monte Carlo is demonstrated in Figure 14 for the four portfolios leading to the largest contribution at the 3rd portfolio level according to Figure 11 (bottom panel).

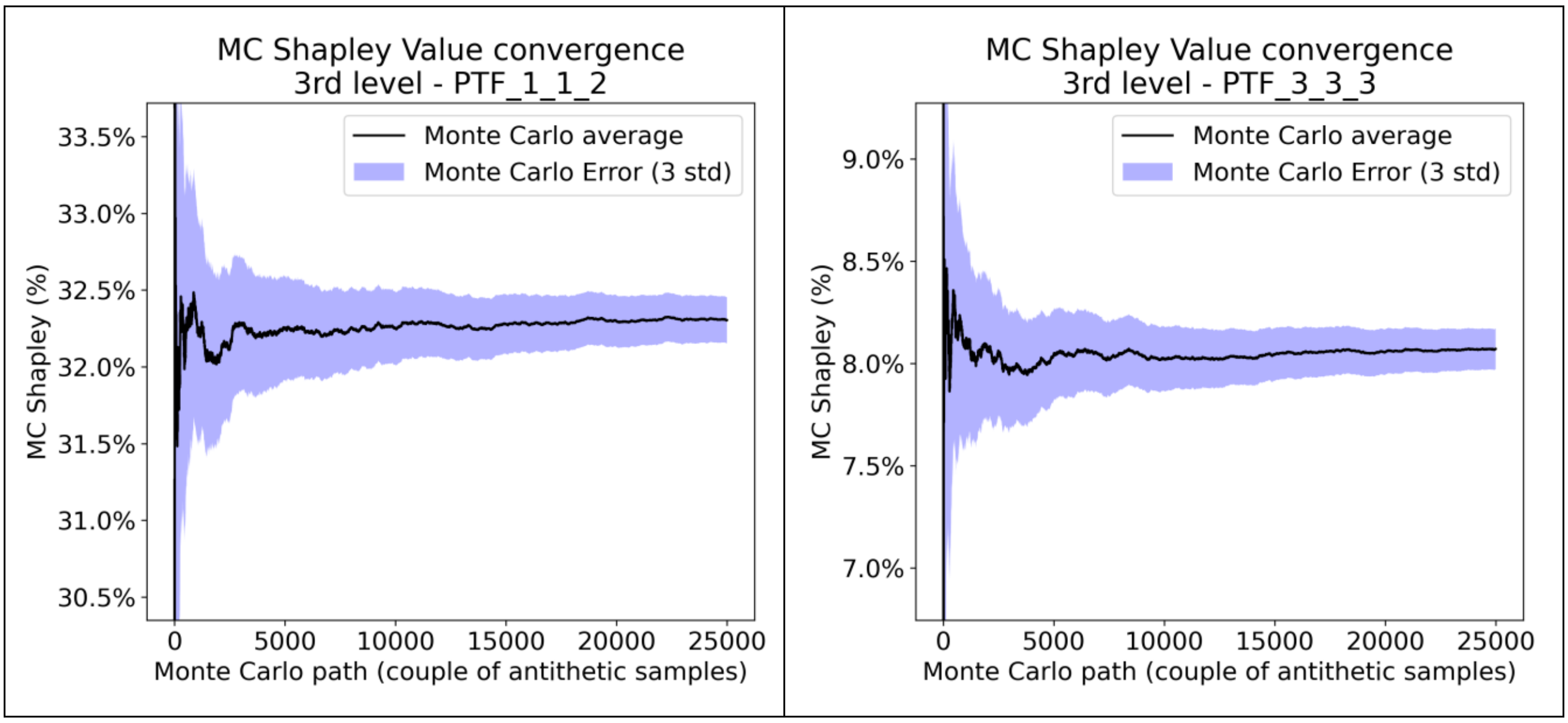

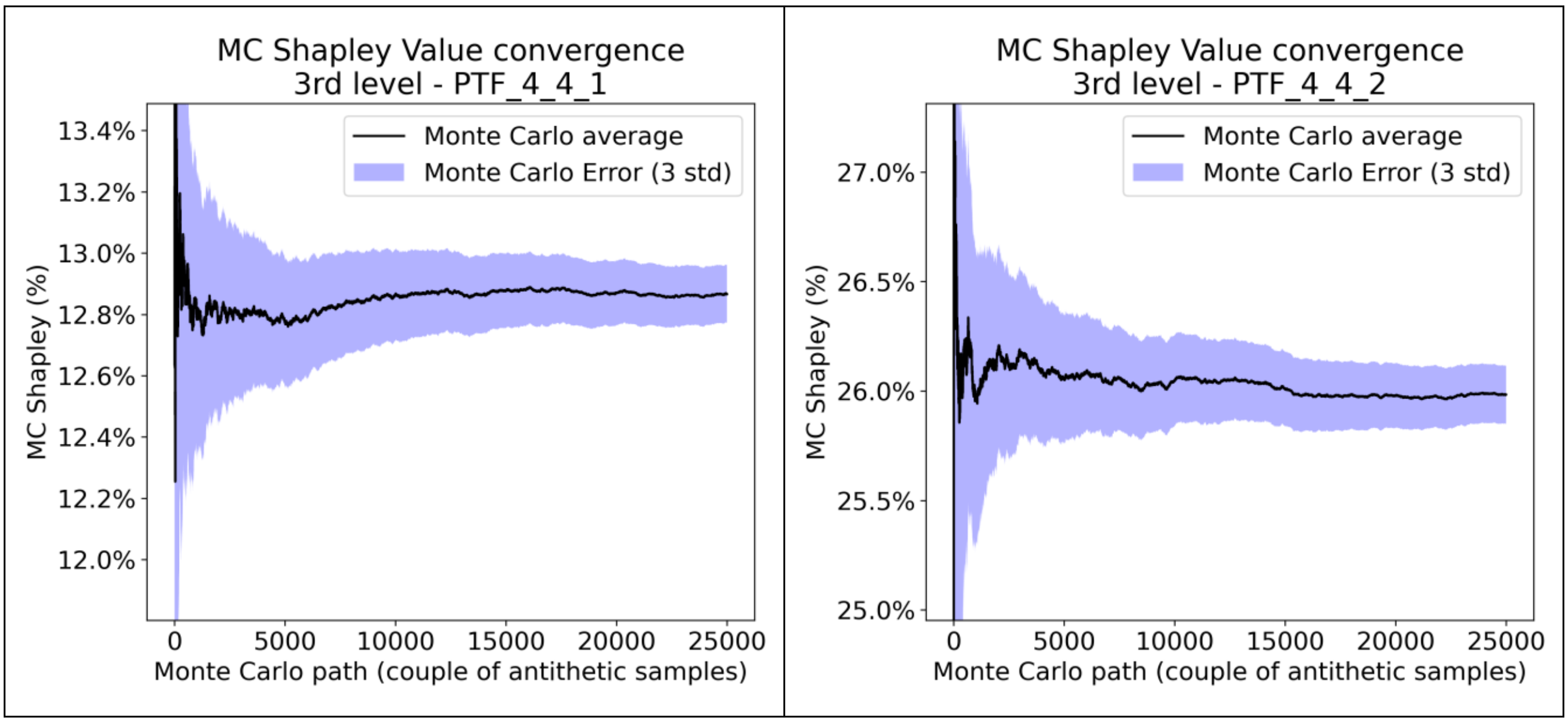

***Figure 14**: Convergence of the Shapley Monte Carlo estimates normalized to total VaR+sVaR $\rho(X)$ in Eq. (71).*

## 5.2. Market Risk Capital Allocation Under FRTB

The Fundamental Review of the Trading Book (CRR3 (2024)) introduces the Standard Approach (SA), based on prescribed rules, and the Internal Model Approach (IMA), based on more sophisticated internal risk models, both at trading desk level. While the IMA is typically adopted by a few large financial institutions, the SA is mandatory for all financial institutions. We refer to App. 8.5 for details on the FRTB SA capital charge formula. In this section we deal with the application of the Shapley allocation strategy to a specific component of the SA risk measure, i.e. the Sensitivity Based Approach (SBA). For the sake of simplicity, we focus on the delta equity risk, since the application of the Shapley allocation to the other risk classes is similar. The Delta ER SBA measure is given by

$$\rho(X) = \mathrm{SBA}^{\mathrm{Delta,\,EQ}}(X) = \sqrt{\sum_{b=1}^{13} K_b^2(X) + 2\sum_{b=1}^{13}\sum_{c=b+1}^{13} \gamma_{b,c}\, S_b(X) S_c(X)}, \tag{72}$$

where we highlighted the dependence of the risk measure on the portfolio $X$, and all terms are explained in App. 8.5.

We apply the Shapley allocation strategy to the same trading book considered in the previous Sec. 5.1, but restricted to the equity positions organized in two hierarchical levels, as depicted in App. 8.5. These equity portfolios contain a broad range of equity financial instruments, stemming from linear (stocks, Futures) and non-linear instruments (options), including exotics. Their delta sensitivities are reported in Figure 16 in App. 8.5, where we notice that PTF_3 hedges the other portfolios and is the most relevant one in terms of absolute equity delta since its delta exposures are higher but partially offset each other.

This portfolio risk allocation is illustrated in Figure 15, displaying the biggest attribution to PTF_3 for the first hierarchical level (top panel), mostly driven by the underlying PTF_3_3 (bottom panel). Again, we observe that, at both levels, Shapley reduces the dominant allocations, redistributing part of the capital charge to the other trading desks. We also observe that, similarly to what observed in the previous section, Proportional Top-Down and Bottom-Up Shapley allocations are aligned with Shapley, thanks to the absence of negative allocations.

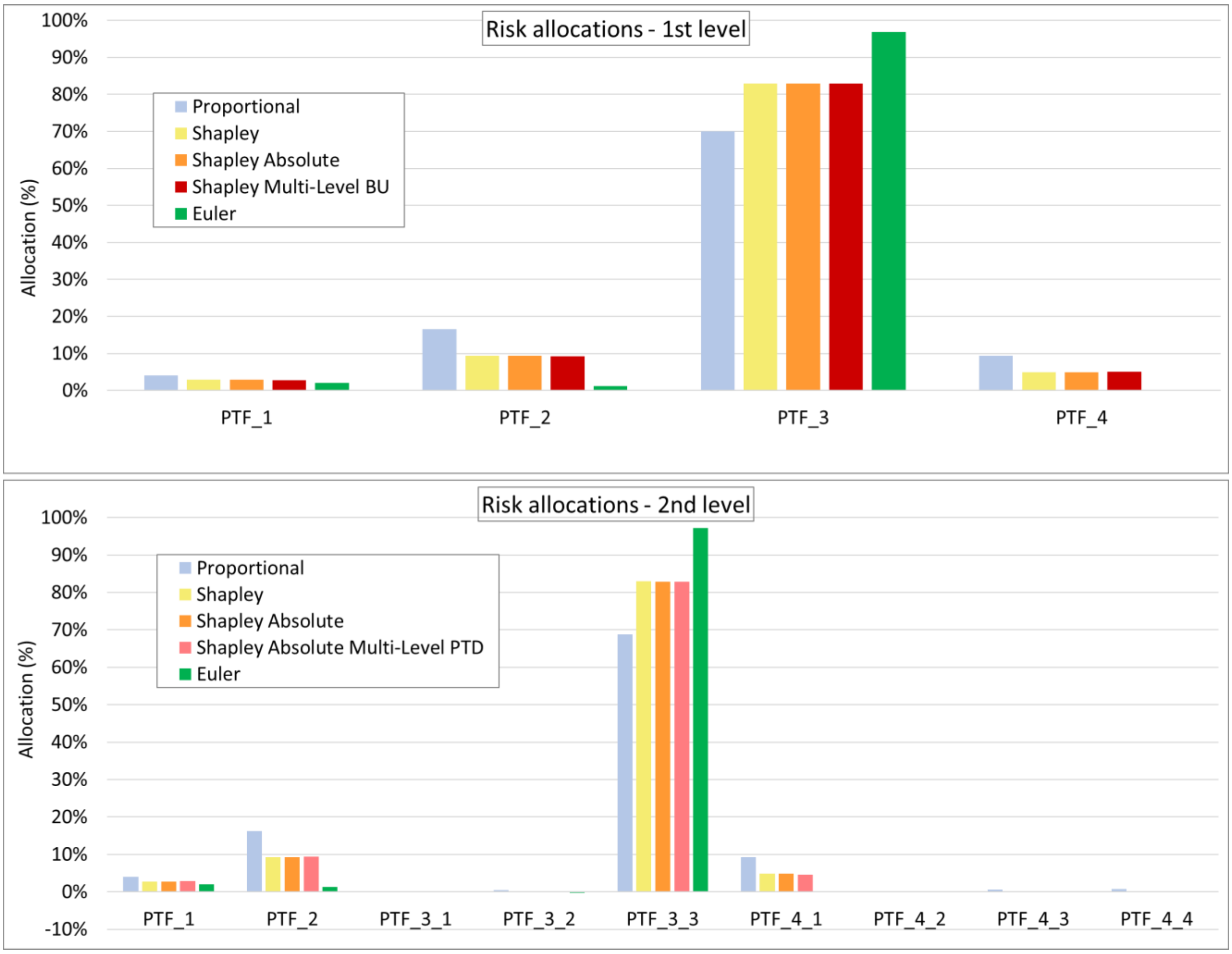

***Figure 15**: Allocations for the first (left hand panel) and second (right hand panel) portfolio hierarchical levels (see Table 9). Shapley and Shapley Absolute coincide for 1st level and almost coincide for 2nd level (PTF_0_2 is allocated -0.1%). All the amounts are normalized to total* $\mathrm{SBA}^{\mathrm{Delta,\,EQ}}\ \rho(X)$ *in Eq.* (72).

# 6. Conclusions

This study provides a systematic investigation of risk allocation strategies, with a focus on theoretical foundations and practical implications in large financial institutions, characterized by a multi-level organisational structure including many business units, and on regulatory market risk measures lacking desirable theoretical properties (e.g. historical VaR).

Our main finding is that the Shapley allocation strategy offers a viable compromise between simplicity, mathematical properties, risk representation and computational cost, provided that, for increasing number of business units, one switches from the analytical allocation to the Monte Carlo allocation. Furthermore, we develop two enhancements to address longstanding challenges in capital attribution: robust procedures to handle negative risk allocations without compromising the full allocation property; and a multi-level allocation strategy consistent with hierarchical organizational structures typical of financial institutions.

Our results have been demonstrated using a specific testing framework, including both simplified settings and realistic financial portfolios under Basel 2.5 and Fundamental Review of the Trading Book (FRTB) regulatory frameworks. These contributions represent a step forward compared to the most recent practice described in Li et al (2016), Schulze (2018), and Li and Xing (2019).

While our empirical applications focus on market risk, our methodological framework is fully general and applicable, in principle, to any financial context where a global risk measure must be decomposed into the contributions of individual components, including - among others - valuation risk, liquidity risk, credit risk, and counterparty credit risk.

# 8. Appendices

## 8.1. Risk Measures

We report below a (non-exhaustive) list of non-additive risk measures which are typically encountered in financial risk management.

- **Valuation risk measures**: Valuation Adjustments (a.k.a. XVAs), taking into account counterparty default risk (CVA/DVA), variation margin and initial margin funding risk (FVA/MVA), capital risk (KVA) (see e.g. Gregory (2020) and Brigo et al. (2013)); Additional Valuation Adjustments (AVAs) taking into account valuation uncertainty of fair valued positions according to the European Capital Requirements Regulation (CRR1 (2013) and CRR2 (2019), see e.g. Bianchetti and Cherubini (2016)); P&L attribution for a portfolio dependent on a set of possibly non-orthogonal risk factors (see e.g. Hagan et al. ); etc.

- **Market risk measures**: Value at Risk (VaR), Stressed Value at Risk (sVaR), Incremental Risk Charge (IRC) (CRR1 (2013) and CRR2 (2019)), and their alter-egos under the Fundamental Review of the Trading Book (FRTB): Expected Shortfall (ES), Default Risk Charge (DRC), Sensitivity-Based Approach (SBA) (CRR3 (2024)).

- **Liquidity risk measures**: Liquidity at Risk (LaR), Liquidity Coverage Ratio (LCR), Net Stable Funding Ratio (NSFR), etc. (CRR1 (2013) and CRR2 (2019)).

- **Credit risk measures**: Exposure at Default (EAD), credit default risk, concentration risk, country risk, etc. (CRR1 (2013) and CRR2 (2019)).

- **Counterparty credit risk measures:** Expected Positive Exposure (EPE), Effective EPE (EEPE), Potential Future Exposure (PFE), Effective PFE (EPFE), etc. (CRR1 (2013) and CRR2 (2019)).

## 8.2. Proof of Full Allocation Property for Shapley Monte Carlo

We prove that the full allocation property is satisfied for each Monte Carlo sample in Shapley Monte Carlo allocation discussed in Sec. 3.1., i.e.

$$\sum_{i=1}^{n} K_{i,m} = K = \rho(X), \forall m \in [1, N_{MC}] \subset \mathbb{N}, 1 \leq m \leq N_{MC}. \tag{73}$$

As shown in Algorithm 1, step 2.4.3, for each MC sample $\pi$ each component of the sum of Eq. (73) is made of the sum of the old and the new risk measure computed on a different coalition, as shown in the following Table 4. For simplicity, we do not consider the antithetic variable of Algorithm 1: the proof still applies, and the generalization is straightforward.

From Table 4 we observe that the new component of i-th risk unit is also the old one of i+1th risk unit.

| $i$ | $\rho_{old}$ | $\rho_{new}$ |
|---|---|---|
| 1 | 0 | $\rho\left(X_{P^{\pi}_{\alpha_1}} + X_1\right)$ |
| 2 | $\rho\left(X_{P^{\pi}_{\alpha_1}} + X_1\right)$ | $\rho\left(X_{P^{\pi}_{\alpha_2}} + X_2\right)$ |
| ... | | |
| $n-1$ | $\rho\left(X_{P^{\pi}_{\alpha_{n-2}}} + X_{n-2}\right)$ | $\rho\left(X_{P^{\pi}_{\alpha_{n-1}}} + X_{n-1}\right)$ |
| $n$ | $\rho\left(X_{P^{\pi}_{\alpha_{n-1}}} + X_{n-1}\right)$ | $\rho\left(X_{P^{\pi}_{\alpha_n}} + X_n\right)$ |

***Table 4**: new and old components for each risk unit involved in the sum of Eq. (73) of Algorithm 1. For each row, we introduced the following $\alpha_i = \pi^{-1}(i)$.*

Hence, the sum in Eq. (73) is telescopic, and therefore all the terms but the first (which is null) and the last one cancel out,

$$\sum_{i=1}^{n} K_{i,m} = \rho\left(X_{P^{\pi}_{\alpha_n}} + X_n\right), \tag{74}$$

where

$$P^{\pi}_{\alpha_n} = \{j \in \mathcal{N} \mid \pi(j) < \pi(\pi^{-1}(n))\} = \{j \in \mathcal{N} \mid \pi(j) < n\} = \{1,2,\dots,n-1\} = \mathcal{N}/\{n\} \tag{75}$$

and therefore

$$\rho\left(X_{P^{\pi}_{\pi^{-1}(n)}} + X_n\right) = \rho\left(X_{\mathcal{N}/\{n\}} + X_n\right) = \rho(X_{\mathcal{N}}) = K. \tag{76}$$

From Eq. (74) and Eq. (76) we obtain Eq. (73).

## 8.3. Illustrative Example of Algorithm 1

We provide in the following an illustrative example of Algorithm 1. Let's assume that $\mathcal{N} = \{1,2,3,4,5\}$, and hence $n = 5$. We initialise $[K_{1,0}, \dots, K_{n,0}] = [0, \dots, 0]$ (step 1) and we show below the steps for the first MC path of the loop of step 2, namely $m = 1$.

We set $\rho_{old} = 0$ (step 2.1) and we assume that the permutation sample is (step 2.2) $\pi = \{2,4,5,3,1\}$. The corresponding antithetic permutation sample is therefore (step 2.3) $\hat{\pi} = \{1,3,5,4,2\}$. We now perform the iteration of step 2.4.

- $i = 1$: $\alpha_1 := \pi^{-1}(1) = 2$ and $\hat{\alpha}_1 \coloneqq \hat{\pi}^{-1}(1) = 1$. Therefore, $P^{\pi}_{\alpha_1} = P^{\pi}_2 = \emptyset$ and $P^{\hat{\pi}}_{\hat{\alpha}_1} = P^{\hat{\pi}}_1 = \emptyset$, hence (from Eq. (5)) we set $\rho_{new} = \frac{1}{2}[\rho(X_{\emptyset} + X_2) + \rho(X_{\emptyset} + X_1)] = \frac{1}{2}[\rho(X_2) + \rho(X_1)]$ . We therefore update the Monte Carlo estimate of portfolio $\alpha_1 = 2$, that is $K_{\alpha,m} = K_{2,1} = \rho_{new}$, also computing $K^2_{2,1}$. Finally, we assign $\rho_{old} = \frac{1}{2}[\rho(X_2) + \rho(X_1)]$.
- $i = 2$: $\alpha_2 = 4$ and $\hat{\alpha}_1 = 3$. Therefore, $P^{\pi}_{\alpha_2} = P^{\pi}_4 = \{2\}$, and $P^{\hat{\pi}}_{\hat{\alpha}_2} = P^{\hat{\pi}}_3 = \{1\}$, hence we update $\rho_{new} = \frac{1}{2}\left[\rho\left(X_{\{2,4\}}\right) + \rho\left(X_{\{1,3\}}\right)\right]$. We therefore update the Monte Carlo estimate of portfolio $\alpha_2 = 4$, that is $K_{4,1} = \rho_{new} - \rho_{old}$, also computing $K^2_{4,1}$. Finally, we update $\rho_{old} = \rho_{new}$.
- We continue up to $i = 5$ in the same way.

## 8.4. Basel 2.5

The Basel 2.5 capital charge formula reads as follows (see CRR1 (2013) and CRR2 (2019))

$$\rho(X, X') = -[\mathrm{VaR}(X, 1\%) + \mathrm{sVaR}(X', 1\%)] + \mathrm{IRC}(X, 99.9\%) \tag{77}$$

where both VaR and sVaR rely on 250 daily historical P&L scenarios. The variable $X'$ is introduced since sVaR is calculated on a historical P&L strip corresponding to a stressed time window, different from the one used for VaR, but applies to the same portfolios[8]. We report in the following Table 5 the details on the trading book behind the numerical results provided in Sec. 5.1.

| **Risk Measure** | **VaR + sVaR** | | |
|---|---|---|---|
| Hierarchical level | 1st | 2nd | 3rd |
| Portfolios | PTF_1 | PTF_1_1 | PTF_1_1_1 |
| | | | PTF_1_1_2 |
| | | PTF_1_2 | PTF_1_2_1 |
| | | | PTF_1_2_2 |
| | PTF_2 | PTF_2 | PTF_2 |
| | PTF_3 | PTF_3_1 | PTF_3_1_1 |
| | | | PTF_3_1_2 |
| | | | PTF_3_1_3 |
| | | | PTF_3_1_4 |
| | | PTF_3_2 | PTF_3_2_1 |
| | | | PTF_3_2_2 |
| | | | PTF_3_2_3 |
| | | PTF_3_3 | PTF_3_3_1 |
| | | | PTF_3_3_2 |
| | | | PTF_3_3_3 |
| | | PTF_3_4 | PTF_3_4 |
| | PTF_4 | PTF_4_1 | PTF_4_1_1 |
| | | | PTF_4_1_2 |
| | | PTF_4_2 | PTF_4_2_1 |
| | | | PTF_4_2_2 |
| | | PTF_4_3 | PTF_4_3 |
| | | PTF_4_4 | PTF_4_4_1 |
| | | | PTF_4_4_2 |
| | | PTF_4_5 | PTF_4_5 |
| Number of Business Units/portfolios ($n$) | 4 | 12 | 24 |
| Number of possible coalitions for each portfolio ($2^{n-1}$) | 8 | 2.048 | 8.388.608 |
| Number of risk measure calculations ($n \times 2^{n-1}$) | 32 | 24.576 | 201.326.592 |
| Shapley Approach | Exact | Exact | Monte Carlo |

[8] The stressed time window is kept fixed in every sVaR computation.

***Table 5**: details on the trading book behind the numerical results provided in Sec. 5.1. We report the risk measures considered, the business units (portfolios) for three distinct hierarchical levels, and some figures on the computational effort involved in the Shapley allocation strategy.*

We show below the main figures (Standalone risk measures and correlation matrices of historical scenarios) of the portfolios analysed in Sec. 5.1, for each of the three hierarchical levels of Table 5.

| | | Ptf | | | |
|---|---|---|---|---|---|
| Ptf | VaR | 1 | 2 | 3 | 4 |
| 1 | 7% | 100 | 0 | -55 | -28 |
| 2 | 1% | 0 | 100 | 7 | -1 |
| 3 | 17% | -55 | 7 | 100 | 40 |
| 4 | 18% | -28 | -1 | 40 | 100 |

| | | Ptf | | | |
|---|---|---|---|---|---|
| Ptf | SVaR | 1 | 2 | 3 | 4 |
| 1 | 5% | 100 | -81 | 47 | -40 |
| 2 | 42% | -81 | 100 | -51 | 37 |
| 3 | 72% | 47 | -51 | 100 | -39 |
| 4 | 0% | -40 | 37 | -39 | 100 |

***Table 6**: Value at Risk and Stressed Value at Risk standalone figures normalized to total VaR+sVaR ($\rho(X)$ in Eq. (71)) (in blue) and corresponding historical correlation matrices (values in percentage) of the 1st level portfolios (in red and green for negative and positive correlations, respectively). The intensity of the colours increases proportionally to the measured amount.*

| | | Ptf | | | | | | | | | | | |
|---|---|---|---|---|---|---|---|---|---|---|---|---|---|
| Ptf | VaR | 1_1 | 1_2 | 2 | 3_1 | 3_2 | 3_3 | 3_4 | 4_1 | 4_2 | 4_3 | 4_4 | 4_5 |
| 1_1 | 6% | 100 | 62 | 2 | -51 | -17 | -41 | 39 | -27 | -24 | 3 | -15 | 25 |
| 1_2 | 1% | 62 | 100 | -8 | -73 | -36 | -43 | -10 | -11 | -24 | -48 | -42 | 37 |
| 2 | 1% | 2 | -8 | 100 | 12 | 23 | -2 | 16 | 7 | -2 | 10 | -3 | 8 |
| 3_1 | 8% | -51 | -73 | 12 | 100 | 41 | 55 | 22 | 21 | 28 | 53 | 47 | -40 |
| 3_2 | 2% | -17 | -36 | 23 | 41 | 100 | 22 | 33 | 7 | 25 | 24 | -4 | 28 |
| 3_3 | 11% | -41 | -43 | -2 | 55 | 22 | 100 | 1 | 13 | 18 | 32 | 18 | -26 |
| 3_4 | 0% | 39 | -10 | 16 | 22 | 33 | 1 | 100 | -24 | 8 | 61 | 30 | 9 |
| 4_1 | 4% | -27 | -11 | 7 | 21 | 7 | 13 | -24 | 100 | 13 | -26 | 5 | -2 |
| 4_2 | 5% | -24 | -24 | -2 | 28 | 25 | 18 | 8 | 13 | 100 | -2 | 17 | 5 |
| 4_3 | 1% | 3 | -48 | 10 | 53 | 24 | 32 | 61 | -26 | -2 | 100 | 54 | -36 |
| 4_4 | 17% | -15 | -42 | -3 | 47 | -4 | 18 | 30 | 5 | 17 | 54 | 100 | -46 |
| 4_5 | 1% | 25 | 37 | 8 | -40 | 28 | -26 | 9 | -2 | 5 | -36 | -46 | 100 |

| | | Ptf | | | | | | | | | | | |
|---|---|---|---|---|---|---|---|---|---|---|---|---|---|
| Ptf | SVaR | 1_1 | 1_2 | 2 | 3_1 | 3_2 | 3_3 | 3_4 | 4_1 | 4_2 | 4_3 | 4_4 | 4_5 |
| 1_1 | 56% | 100 | -8 | -82 | 35 | 75 | 16 | 79 | -79 | -41 | 78 | -33 | 78 |
| 1_2 | 2% | -8 | 100 | 15 | -46 | -17 | -28 | -2 | 20 | 21 | -20 | 17 | -8 |
| 2 | 5% | -82 | 15 | 100 | -32 | -67 | -25 | -75 | 89 | 35 | -77 | 28 | -55 |
| 3_1 | 8% | 35 | -46 | -32 | 100 | 41 | 28 | 36 | -34 | -23 | 64 | -17 | 26 |
| 3_2 | 20% | 75 | -17 | -67 | 41 | 100 | 28 | 71 | -68 | -38 | 68 | -32 | 66 |
| 3_3 | 29% | 16 | -28 | -25 | 28 | 28 | 100 | 23 | -17 | 1 | 25 | -32 | 17 |
| 3_4 | 2% | 79 | -2 | -75 | 36 | 71 | 23 | 100 | -71 | -24 | 76 | -20 | 51 |
| 4_1 | 9% | -79 | 20 | 89 | -34 | -68 | -17 | -71 | 100 | 47 | -78 | 32 | -55 |
| 4_2 | 17% | -41 | 21 | 35 | -23 | -38 | 1 | -24 | 47 | 100 | -49 | 41 | -18 |
| 4_3 | 3% | 78 | -20 | -77 | 64 | 68 | 25 | 76 | -78 | -49 | 100 | -35 | 50 |
| 4_4 | 75% | -33 | 17 | 28 | -17 | -32 | -32 | -20 | 32 | 41 | -35 | 100 | -28 |
| 4_5 | 4% | 78 | -8 | -55 | 26 | 66 | 17 | 51 | -55 | -18 | 50 | -28 | 100 |

***Table 7**: same as in Table 6 for 2nd level portfolios.*

| Ptf | VaR | 1_1_1 | 1_1_2 | 1_2_1 | 1_2_2 | 2 | 3_1_1 | 3_1_2 | 3_1_3 | 3_1_4 | 3_2_1 | 3_2_2 | 3_2_3 | 3_3_1 | 3_3_2 | 3_3_3 | 3_4 | 4_1_1 | 4_1_2 | 4_2_1 | 4_2_2 | 4_3 | 4_4_1 | 4_4_2 | 4_5 |
|---|---|---|---|---|---|---|---|---|---|---|---|---|---|---|---|---|---|---|---|---|---|---|---|---|---|
| 1_1_1 | 10% | 100 | -74 | -61 | 98 | -10 | -9 | -61 | -25 | -96 | -79 | 12 | -34 | 21 | -12 | -61 | -21 | -11 | -13 | -25 | -30 | -53 | -76 | -33 | 38 |
| 1_1_2 | 8% | -74 | 100 | 44 | -67 | 15 | -2 | 59 | 10 | 75 | 56 | 23 | 6 | -32 | -28 | 56 | 61 | -7 | 4 | 14 | 31 | 72 | 85 | 33 | -28 |
| 1_2_1 | 1% | -61 | 44 | 100 | -59 | 6 | 4 | 40 | 17 | 57 | 47 | -2 | 19 | -13 | 12 | 33 | 22 | 11 | 3 | 19 | 23 | 32 | 45 | 23 | -18 |
| 1_2_2 | 2% | 98 | -67 | -59 | 100 | -9 | -6 | -55 | -28 | -91 | -74 | 9 | -32 | 22 | -10 | -61 | -18 | -8 | -13 | -24 | -33 | -53 | -69 | -33 | 38 |
| 2 | 1% | -10 | 15 | 6 | -9 | 100 | 11 | 5 | 7 | 11 | 19 | 5 | 18 | -12 | -4 | 6 | 16 | 7 | 4 | -2 | -1 | 10 | 7 | -6 | 8 |
| 3_1_1 | 2% | -9 | -2 | 4 | -6 | 11 | 100 | 5 | 2 | 8 | 15 | -9 | 9 | 18 | 20 | 6 | -2 | 12 | 15 | 5 | 0 | -1 | 5 | 12 | -2 |
| 3_1_2 | 3% | -61 | 59 | 40 | -55 | 5 | 5 | 100 | 7 | 67 | 50 | -6 | 20 | -8 | -1 | 54 | 29 | 9 | 6 | 3 | 9 | 57 | 72 | 22 | -39 |
| 3_1_3 | 4% | -25 | 10 | 17 | -28 | 7 | 2 | 7 | 100 | 24 | 22 | -2 | 16 | -4 | 12 | 19 | 1 | 7 | 5 | 17 | 9 | 4 | 10 | -2 | -6 |
| 3_1_4 | 4% | -96 | 75 | 57 | -91 | 11 | 8 | 67 | 24 | 100 | 81 | -17 | 39 | -19 | 14 | 61 | 19 | 16 | 13 | 32 | 20 | 53 | 83 | 30 | -39 |
| 3_2_1 | 1% | -79 | 56 | 47 | -74 | 19 | 15 | 50 | 22 | 81 | 100 | -26 | 69 | -14 | 25 | 49 | 16 | 23 | 17 | 27 | 13 | 34 | 66 | 0 | -8 |
| 3_2_2 | 1% | 12 | 23 | -2 | 9 | 5 | -9 | -6 | -2 | -17 | -26 | 100 | -36 | -23 | -48 | 4 | 50 | -30 | -14 | -20 | 44 | 27 | -9 | 13 | 22 |
| 3_2_3 | 2% | -34 | 6 | 19 | -32 | 18 | 9 | 20 | 16 | 39 | 69 | -36 | 100 | 2 | 44 | 20 | -13 | 28 | 11 | 36 | -12 | -11 | 27 | -37 | 21 |
| 3_3_1 | 5% | 21 | -32 | -13 | 22 | -12 | 18 | -8 | -4 | -19 | -14 | -23 | 2 | 100 | 25 | -14 | -28 | -1 | 12 | 4 | -17 | -22 | -18 | -8 | -4 |
| 3_3_2 | 4% | -12 | -28 | 12 | -10 | -4 | 20 | -1 | 12 | 14 | 25 | -48 | 44 | 25 | 100 | -19 | -47 | 40 | 17 | 31 | -23 | -50 | -6 | -18 | 8 |
| 3_3_3 | 9% | -61 | 56 | 33 | -61 | 6 | 6 | 54 | 19 | 61 | 49 | 4 | 20 | -14 | -19 | 100 | 31 | 2 | -4 | 8 | 19 | 62 | 64 | 15 | -29 |
| 3_4 | 0% | -21 | 61 | 22 | -18 | 16 | -2 | 29 | 1 | 19 | 16 | 50 | -13 | -28 | -47 | 31 | 100 | -26 | -7 | 1 | 35 | 61 | 42 | 20 | 9 |
| 4_1_1 | 3% | -11 | -7 | 11 | -8 | 7 | 12 | 9 | 7 | 16 | 23 | -30 | 28 | -1 | 40 | 2 | -26 | 100 | 10 | 22 | -16 | -38 | 6 | -8 | -1 |
| 4_1_2 | 2% | -13 | 4 | 3 | -13 | 4 | 15 | 6 | 5 | 13 | 17 | -14 | 11 | 12 | 17 | -4 | -7 | 10 | 100 | -2 | -5 | 7 | 6 | 17 | -2 |
| 4_2_1 | 5% | -25 | 14 | 19 | -24 | -2 | 5 | 3 | 17 | 32 | 27 | -20 | 36 | 4 | 31 | 8 | 1 | 22 | -2 | 100 | -10 | -8 | 31 | 6 | 5 |
| 4_2_2 | 1% | -30 | 31 | 23 | -33 | -1 | 0 | 9 | 9 | 20 | 13 | 44 | -12 | -17 | -23 | 19 | 35 | -16 | -5 | -10 | 100 | 31 | 8 | 14 | 2 |
| 4_3 | 1% | -53 | 72 | 32 | -53 | 10 | -1 | 57 | 4 | 53 | 34 | 27 | -11 | -22 | -50 | 62 | 61 | -38 | 7 | -8 | 31 | 100 | 64 | 40 | -36 |
| 4_4_1 | 5% | -76 | 85 | 45 | -69 | 7 | 5 | 72 | 10 | 83 | 66 | -9 | 27 | -18 | -6 | 64 | 42 | 6 | 6 | 31 | 8 | 64 | 100 | 31 | -37 |
| 4_4_2 | 16% | -33 | 33 | 23 | -33 | -6 | 12 | 22 | -2 | 30 | 0 | 13 | -37 | -8 | -18 | 15 | 20 | -8 | 17 | 6 | 14 | 40 | 31 | 100 | -40 |
| 4_5 | 1% | 38 | -28 | -18 | 38 | 8 | -2 | -39 | -6 | -39 | -8 | 22 | 21 | -4 | 8 | -29 | 9 | -1 | -2 | 5 | 2 | -36 | -37 | -40 | 100 |

| Ptf | SVaR | 1_1_1 | 1_1_2 | 1_2_1 | 1_2_2 | 2 | 3_1_1 | 3_1_2 | 3_1_3 | 3_1_4 | 3_2_1 | 3_2_2 | 3_2_3 | 3_3_1 | 3_3_2 | 3_3_3 | 3_4 | 4_1_1 | 4_1_2 | 4_2_1 | 4_2_2 | 4_3 | 4_4_1 | 4_4_2 | 4_5 |
|---|---|---|---|---|---|---|---|---|---|---|---|---|---|---|---|---|---|---|---|---|---|---|---|---|---|
| 1_1_1 | 10% | 100 | -45 | -2 | 88 | 36 | 2 | -52 | 16 | -63 | -54 | -36 | -3 | 18 | 7 | -55 | -33 | 32 | 40 | 30 | -4 | -47 | -39 | 29 | -25 |
| 1_1_2 | 60% | -45 | 100 | -8 | -16 | -82 | -27 | 43 | -19 | 51 | 62 | 86 | -21 | -43 | -51 | 62 | 79 | -63 | -80 | -45 | 20 | 81 | 37 | -44 | 77 |
| 1_2_1 | 1% | -2 | -8 | 100 | 7 | 12 | 2 | -22 | 8 | -12 | 2 | -10 | 7 | 0 | 3 | -9 | 1 | 3 | 15 | 15 | -12 | -10 | 7 | 12 | -8 |
| 1_2_2 | 1% | 88 | -16 | 7 | 100 | 11 | -6 | -39 | 11 | -43 | -31 | -10 | -9 | 4 | -8 | -34 | -3 | 13 | 17 | 17 | -1 | -18 | -22 | 18 | -5 |
| 2 | 5% | 36 | -82 | 12 | 11 | 100 | 35 | -39 | 20 | -43 | -50 | -75 | 18 | 34 | 64 | -65 | -75 | 71 | 89 | 36 | -14 | -77 | -42 | 39 | -55 |
| 3_1_1 | 3% | 2 | -27 | 2 | -6 | 35 | 100 | -9 | -5 | -8 | -19 | -22 | 4 | 14 | 27 | -22 | -31 | 26 | 39 | 14 | -2 | -27 | -13 | 8 | -17 |
| 3_1_2 | 5% | -52 | 43 | -22 | -39 | -39 | -9 | 100 | -15 | 47 | 37 | 41 | 12 | -16 | -22 | 50 | 41 | -31 | -42 | -26 | 19 | 54 | 41 | -23 | 29 |
| 3_1_3 | 4% | 16 | -19 | 8 | 11 | 20 | -5 | -15 | 100 | -13 | -7 | -25 | 6 | 9 | -1 | -21 | -8 | 17 | 23 | 24 | -11 | -19 | -2 | 28 | -13 |
| 3_1_4 | 6% | -63 | 51 | -12 | -43 | -43 | -8 | 47 | -13 | 100 | 49 | 48 | -12 | -21 | -14 | 49 | 41 | -34 | -47 | -40 | 37 | 76 | 21 | -37 | 32 |
| 3_2_1 | 2% | -54 | 62 | 2 | -31 | -50 | -19 | 37 | -7 | 49 | 100 | 54 | 21 | -21 | -19 | 51 | 67 | -45 | -49 | -26 | 8 | 60 | 53 | -30 | 49 |
| 3_2_2 | 19% | -36 | 86 | -10 | -10 | -75 | -22 | 41 | -25 | 48 | 54 | 100 | -30 | -45 | -46 | 59 | 74 | -58 | -78 | -61 | 32 | 78 | 34 | -51 | 57 |
| 3_2_3 | 9% | -3 | -21 | 7 | -9 | 18 | 4 | 12 | 6 | -12 | 21 | -30 | 100 | 19 | 30 | 4 | -10 | 15 | 18 | 37 | -17 | -23 | 21 | 11 | 13 |
| 3_3_1 | 20% | 18 | -43 | 0 | 4 | 34 | 14 | -16 | 9 | -21 | -21 | -45 | 19 | 100 | 22 | -19 | -29 | 34 | 41 | 36 | -12 | -37 | -9 | 24 | -30 |
| 3_3_2 | 6% | 7 | -51 | 3 | -8 | 64 | 27 | -22 | -1 | -14 | -19 | -46 | 30 | 22 | 100 | -26 | -50 | 40 | 57 | 22 | -2 | -51 | -25 | -16 | -24 |
| 3_3_3 | 23% | -55 | 62 | -9 | -34 | -65 | -22 | 50 | -21 | 49 | 51 | 59 | 4 | -19 | -26 | 100 | 56 | -39 | -63 | -31 | 19 | 64 | 44 | -61 | 45 |
| 3_4 | 2% | -33 | 79 | 1 | -3 | -75 | -31 | 41 | -8 | 41 | 67 | 74 | -10 | -29 | -50 | 56 | 100 | -60 | -68 | -26 | 14 | 76 | 59 | -35 | 51 |
| 4_1_1 | 3% | 32 | -63 | 3 | 13 | 71 | 26 | -31 | 17 | -34 | -45 | -58 | 15 | 34 | 40 | -39 | -60 | 100 | 68 | 34 | -13 | -61 | -32 | 35 | -43 |
| 4_1_2 | 7% | 40 | -80 | 15 | 17 | 89 | 39 | -42 | 23 | -47 | -49 | -78 | 18 | 41 | 57 | -63 | -68 | 68 | 100 | 51 | -21 | -77 | -34 | 41 | -55 |
| 4_2_1 | 17% | 30 | -45 | 15 | 17 | 36 | 14 | -26 | 24 | -40 | -26 | -61 | 37 | 36 | 22 | -31 | -26 | 34 | 51 | 100 | -28 | -53 | 9 | 45 | -21 |
| 4_2_2 | 5% | -4 | 20 | -12 | -1 | -14 | -2 | 19 | -11 | 37 | 8 | 32 | -17 | -12 | -2 | 19 | 14 | -13 | -21 | -28 | 100 | 31 | -15 | -36 | 18 |
| 4_3 | 3% | -47 | 81 | -10 | -18 | -77 | -27 | 54 | -19 | 76 | 60 | 78 | -23 | -37 | -51 | 64 | 76 | -61 | -77 | -53 | 31 | 100 | 37 | -45 | 50 |
| 4_4_1 | 22% | -39 | 37 | 7 | -22 | -42 | -13 | 41 | -2 | 21 | 53 | 34 | 21 | -9 | -25 | 44 | 59 | -32 | -34 | 9 | -15 | 37 | 100 | -9 | 17 |
| 4_4_2 | 73% | 29 | -44 | 12 | 18 | 39 | 8 | -23 | 28 | -37 | -30 | -51 | 11 | 24 | -16 | -61 | -35 | 35 | 41 | 45 | -36 | -45 | -9 | 100 | -33 |
| 4_5 | 4% | -25 | 77 | -8 | -5 | -55 | -17 | 29 | -13 | 32 | 49 | 57 | 13 | -30 | -24 | 45 | 51 | -43 | -55 | -21 | 18 | 50 | 17 | -33 | 100 |

***Table 8****: same as in Table 6 for 3rd level portfolios.*

## 8.5. Fundamental Review of Trading Book

The FRTB total capital charge formula (CRR3 (2024)) reads as follows

$$C = \min\left[\mathrm{IMA}_{\mathrm{IMA\ ptf}} + \mathrm{PLAAddOn} + \mathrm{SA}_{\mathrm{SA\ ptf}}; \mathrm{SA}_{\mathrm{IMA\ ptf+SA\ ptf}}\right] + \max\left[0; \mathrm{IMA}_{\mathrm{IMA\ ptf}} - \mathrm{SA}_{\mathrm{SA\ ptf}}\right]. \quad (78)$$

In Eq. (78) the subscripts $\mathrm{IMA\ ptf}$ and $\mathrm{SA\ ptf}$ points out whether the corresponding risk measures are computed on the subset of the trading desks subject either to the Internal Model Approach (IMA) or to the Standard Approach (SA), respectively. IMA and SA risk measures are composed of several risk components as show in eqs. (79) and (80) below,

$$\mathrm{IMA_s} = \mathrm{IMCC_s} + \mathrm{SES_s} + \mathrm{DRC_{IMA,s}}, \quad (79)$$

$$\mathrm{SA_s} = \max_{c}\left[\mathrm{SBA_s^{Delta}}(c) + \mathrm{SBA_s^{Vega}}(c) + \mathrm{SBA_s^{Curvature}}(c)\right] + \mathrm{DRC_{SA,s}} + \mathrm{RRAO_s}, \quad (80)$$

where $\mathrm{s} \in \{\mathrm{IMA\ ptf}, \mathrm{SA\ ptf}\}$, SBA stands for Sensitivity Based Approach, and $c \in \{H, M, L\}$ refer to the parametrization used among three possible choices (denominated as High, Medium, Low). i.e. the one leading to the maximum Delta, Vega and Curvature components. Since in this paper the additional variable $c$ is not considered, it is excluded hereafter. For DRC measure, an additional subscript is included to distinguish between the standard and internal model formulas. For the full definitions of all the quantities involved in eqs. (78), (79) and (80), and for the portfolio assignation rules to IMA or SA we refer to CRR3 (2024).

From Eq. (78) we notice that the FRTB capital charge is floored by the SA capital charge computed on the whole perimeter ($\mathrm{IMA\ ptf} + \mathrm{SA\ ptf}$); therefore, the standard approach plays a crucial role for the determination of the final capital charge. In this work we focus, for the sake of simplicity, on $\mathrm{SBA_s^{Delta}}$ risk measure, for which we provide in Eq. (81) below its breakdown on each asset class prescribed by the regulation

$$\mathrm{SBA^{Delta}} = \mathrm{SBA^{Delta,\,EQ}} + \mathrm{SBA^{Delta,GIRR}} + \mathrm{SBA^{Delta,CS}} + \mathrm{SBA^{Delta,COMM}} + \mathrm{SBA^{Delta,FX}}. \quad (81)$$

Each component in Eq. (81) is computed by aggregation of the corresponding delta sensitivities multiplied by specific coefficients. The aggregations are made on different sub-portfolios (a.k.a. "buckets") whose nature depends on the asset class considered. For what concerns the Equity asset class the $\mathrm{SBA^{Delta,\,EQ}}$ risk measure reads as follows

$$\mathrm{SBA^{Delta,\,EQ}} = \sqrt{\sum_{b=1}^{13} K_b^2 + 2\sum_{b=1}^{13}\sum_{c=b+1}^{13} \gamma_{b,c}\, S_b S_c} := \sqrt{S}, \quad (82)$$

where 13 is the number of buckets for the Equity asset class, relative to general characteristics of the underlying equities, $K_b, S_b$ are given by

$$K_b = \begin{cases} \sqrt{\sum_{k=1}^{N_b} WS_{b,k}^2 + 2\sum_{k=1}^{N_b}\sum_{l=k+1}^{N_b} \rho_{k,l} WS_{b,k} WS_{b,l}} & b \neq 11 \\ \sum_{k=1}^{N_b} |WS_{b,k}| & b = 11 \end{cases} \quad (83)$$

$$S_b = \begin{cases} \sum_{k=1}^{N_b} \mathrm{WS}_{b,k} & S \geq 0 \\ \max\left[\min\left[\sum_{k=1}^{N_b} \mathrm{WS}_{b,k}, K_b\right], -K_b\right] & S < 0 \end{cases} \tag{84}$$

$$WS_{b,k} = w_b s_{b,k}, \qquad s_{b,k} = \sum_{i\in b} \frac{\partial V_{i,b}}{\partial u_k}, \tag{85}$$

where $s_{b,k}$ is the total Equity Delta sensitivity of trades in bucket $b$ with respect to risk factors $u_k, k = 1, \dots, N_b$, and $\gamma, \rho, w$ are parameters fixed by the regulation. We refer to CRR3 (2024) for further details.

We report in the following Table 5 the details on the trading book behind the numerical results provided in Sec. 5.2.

| **Risk Measure** | **SBA Delta Equity** | |
|---|---|---|
| Hierarchical level | 1st | 2nd |
| | PTF_1 | PTF_1 |
| | PTF_2 | PTF_2 |
| | PTF_3 | PTF_3_1 |
| | | PTF_3_2 |
| | | PTF_3_3 |
| | PTF_4 | PTF_4_1 |
| | | PTF_4_2 |
| | | PTF_4_3 |
| | | PTF_4_4 |
| Number of Business Units/portfolios ($n$) | 4 | 9 |
| Number of possible coalitions for each portfolio ($2^{n-1}$) | 8 | 256 |
| Number of risk measure calculations ($n \times 2^{n-1}$) | 32 | 2.304 |
| Shapley Approach | Exact | Exact |

***Table 9**: Details on the trading book behind the numerical results provided in Sec. 5.2. We report the risk measures considered, the business units (portfolios) for two distinct hierarchical levels, and some figures on the computational effort involved in the Shapley allocation strategy.*

We show in Figure 16 the portfolio delta sensitivities. We notice that PTF_3 hedges the other portfolios and is the most relevant one in terms of absolute equity delta, since its delta exposures are higher but partially offset each other.

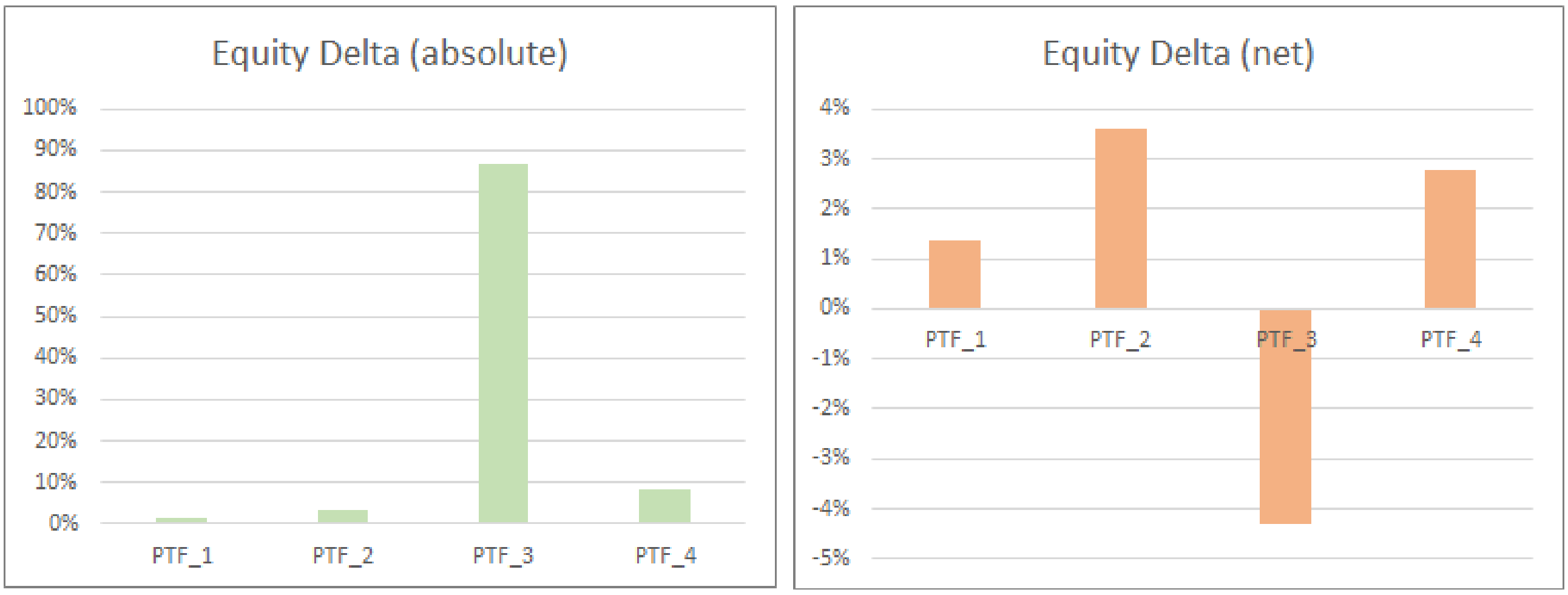

***Figure 16:*** *Portfolio Equity Delta 1% exposure. Left panel: sum of sensitivity values (netted). Right panel: sum of absolute sensitivity values (not netted). All figures normalized to total absolute sensitivity. Portfolio 3 is the leading portfolio in terms of absolute sensitivity.*

## 8.6. Risk Allocation Strategies

We report here a detailed comparison of the mathematical properties of risk allocation strategies discussed in Sec. 2.3.

| | **Standalone** | **Proportional** | **Marginal** | **Shapley** | **Shapley Monte Carlo** | **Euler** | **VaR Euler 1° order** | **VaR Euler 2°order** | **Var Euler Kernel** |
|---|---|---|---|---|---|---|---|---|---|
| **Full allocation** | Not satisfied | Satisfied (ex post) | Satisfied (ex post) | Satisfied (by definition) | Satisfied (by definition, in each path) | Satisfied (homogeneous risk measures) | Almost satisfied | Almost satisfied | Almost satisfied |
| **Computational Effort (per measure evaluation)** | Low | Low | Low | High | Medium | Low | Negligible | Negligible | Negligible |
| **Interactions among risk units** | Not considered | Not considered | Partially considered | Fully considered | Fully considered | Fully considered | Fully considered | Fully considered | Fully considered |
| **Potential negative values** | No | No | Yes | Yes | Yes | Yes | Yes | Yes | Yes |
| **Theoretical/numerical issues** | No | No | Denominator can be small | No | Convergence to be verified empirically | Not applicable in discrete case for VaR | Approximation | Approximation | Approximation + model assumptions |

***Table 10**: comparison of risk allocation strategies.*